\documentclass[pdflatex,sn-mathphys-num]{sn-jnl}

\usepackage{graphicx}%
\usepackage{multirow}%
\usepackage{amsmath,amssymb,amsfonts}%
\usepackage{amsthm}%
\usepackage{mathrsfs}%
\usepackage[title]{appendix}%
\usepackage{xcolor}%
\usepackage{textcomp}%
\usepackage{manyfoot}%
\usepackage{booktabs}%
\usepackage{algorithm}%
\usepackage{algorithmicx}%
\usepackage{algpseudocode}%
\usepackage{listings}%
\usepackage{subfig}%
\usepackage{enumitem}%
\usepackage{caption}%

\renewenvironment{table}[1][]%
{\tableorg[#1]%
	\tablebodyfont%
	\renewcommand\footnotetext[2][]{{\removelastskip\vskip3pt%
			\let\tablebodyfont\tablefootnotefont%
			\hskip0pt\if!##1!\else{\smash{$^{##1}$}}\fi##2\par}}%
}{\endtableorg}

\theoremstyle{thmstyleone}%
\newtheorem{theorem}{Theorem}
\newtheorem{proposition}[theorem]{Proposition}%
\newtheorem{corollary}{Corollary}%

\theoremstyle{thmstyletwo}%
\newtheorem{remark}{Remark}%

\theoremstyle{thmstylethree}%
\newtheorem{definition}{Definition}%

\begin{document}

\title[Local balance in global international relations]{Mathematical Modeling of Local Balance in Signed Networks and Its Applications to Global International Analysis}


\author[1]{\fnm{Fernando} \sur{Diaz-Diaz}}\email{fernandodiaz@ifisc.uib-csic.es}

\author[2]{\fnm{Paolo} \sur{Bartesaghi}}\email{paolo.bartesaghi@unimib.it}

\author*[1]{\fnm{Ernesto} \sur{Estrada}}\email{estrada@ifisc.uib-csic.es}

\affil*[1]{\orgdiv{Institute of Cross-Disciplinary Physics and Complex Systems}, \orgname{IFISC (UIB-CSIC)}, \orgaddress{\city{Palma de Mallorca}, \postcode{07122}, \country{Spain}}
}

\affil[2]{\orgdiv{Department of Statistics and Quantitative Methods}, \orgname{University of Milano - Bicocca}, \orgaddress{\street{Via Bicocca degli Arcimboldi 8}, \city{Milano}, \postcode{20126}, \country{Italy}}}



\abstract{Alliances and conflicts in social, political and economic relations can be represented by positive and negative edges in signed networks. A cycle is said to be positive if the product of its edge signs is positive, otherwise it is negative. Then, a signed network is balanced if and only if all its cycles are positive. An index characterizing how much a signed network deviates from being balanced is known as a global balance index. Here we give a step forward in the characterization of signed networks by defining a local balance index, which characterizes how much a given vertex of a signed network contributes to its global balance. We analyze the mathematical foundations and unique structural properties of this index. Then, we apply this index to the study of the evolution of international relations in the globe for the period 1816-2014. In this way we detect and categorize major historic events based on balance fluctuations, helping our understanding towards new mixed approaches to history based on network theory.}

\keywords{Signed networks; structural balance; international
	relations; geopolitical networks; quantitative history.}


\pacs[MSC Classification]{05C22, 05C50, 05C90, 91D35, 91F10}

\maketitle

\section{Introduction}\label{Introduction}

Complex systems are, by definition, networked \cite{estrada2023complex}. Therefore, their complexities demands the use
of a combination of global and local graph-theoretic invariants to
understand their structures and functions \cite{WassermanFaust1994,Estrada_book}.
A particular case of complex system is the representation of conflicting
interactions by means of signed networks \cite{Signed_graphs}. In this case, for instance,
positive relations may refer to friendship, collaboration or alliances,
and negative ties represent enmities, opposite view about a topic,
or conflicts. When the set of vertices of a signed network can be split into two subsets 
such that every edge between nodes within each subset is positive, and negative
edges only interconnecting vertices of the different sets, the system
is called balanced \cite{harary1953,cartwright1956}. A balanced network
does not contain negative cycles-- that is, cycles in which the product of
the edge signs is negative. When a signed network is not balanced,
it is important to know how unbalanced it is, which will give rise
to an index characterizing the degree of balance of the network as a
whole \cite{kirkley2019balance,balance_5,aref2020multilevel,Estrada_2014,Estrada_2019, talaga2023polarization}.
Moreover, in an unbalanced
network not every vertex contributes the same to the global level
of balance of the graph. Thus, it is important to know the local degree
of balance of the vertices. This concept was proposed by Harary \cite{harary1955local}
and further by Cartwright and Harary \cite{cartwright1956}. However,
these authors did not systematically studied any particular index of
the degree of local balance in signed graphs. 

The theory of network balance emerged from the early days of social
network analysis \cite{heider_1,heider_2}, and nowadays has become
an important paradigm for the analysis of sociometric relations ranging
from individuals, to institutions and countries \cite{Estrada_2014, Estrada_2019, kirkley2019balance,balance_5, talaga2023polarization, tian2022spreading}. In particular, balance theory plays an important role in the analysis of international
relations (IR), where signed networks are used to represent countries
and their alliances/conflicts \cite{Network_IR_3, Network_IR_2,Doreian,Network_IR_4,Network_IR_1,harary1961structural, Galam2014a, Galam2014b}.
This represents an alternative approach to the more classical, statistically-oriented,
frameworks of ``quantitative history'' \cite{Gauthier,Qualitative,Quantitative-2},
and ``cliodynamics'' \cite{cliodynamics_1}.
Coalition models for IR have also been proposed to account for local versus global alignment among countries, showing that, for example, the existence of two competing world coalitions leads to a more stable distribution of actors, while a single world leadership enables the emergence of unstable relationships \cite{Galam1996}. These previous studies
on networks in IR (NIR) have focused on detecting global patterns
on the temporal evolution of the world as a whole. However, to connect
NIR with historical narrative it is necessary to zoom in the individual
contribution of every country to the balance/unbalance of the world
in a given historic period. Consequently, here we define a local balance
index based on algebraic graph theory and study several of its mathematical
properties. In addition, we design an approach based on this index
to identify the major events in the world history for the period
between 1816 and 2014. In this way we connect a graph-based quantitative
analysis of NIR with the qualitative, narrative account of history
\cite{Quantitative-2,Tamura}. The current work can be contextualized
in a general framework of quantitative history based on the so-called
``mixed method'', which ``combines quantitative and qualitative
research techniques, methods, approaches, concepts or language into
a single study'' \cite{mixed-1,mixed_methods}. 

The current paper is subdivided into two interrelated parts. In Part
I we motivate, develop, analyze, interpret and compare a new local
balance index. The goals of this part of the paper are: (i) to provide
a motivation about the necessity for the development of a local balance
index; (ii) to define mathematically a local balance index that accounts
for the role of an individual vertex to the global balance in a signed
network; (iii) to analyze several of the mathematical characteristics
of this local balance index remarking its structural interpretability;
(iv) to provide an interpretation of the local balance index in terms
of a dynamical process taking place in signed (social) networks. In
Part II we develop a general methodology for connecting significant
changes in the local balance index of individual countries with historic
events involving those countries at a given time. The goals of this second
part of the paper are: (i) to apply the local balance index to the
study of IR among countries in the world; (ii) to design and test
a general methodology allowing to match the changes (peaks and valleys)
in the time series of countries' local balance with major historic
events involving those countries at this time; (iii) to perform a
classification of major historic events on the basis of the nature
of the change (peaks and valleys) occurring in the local balance;
(iv) to link the identification of peaks and valleys with a narrative
of historic events; (v) to analyze the temporal series of local balance
of individual countries in their respective networks of IR in the
period 1816-2014.

\section{Preliminaries}
A binary undirected signed network is represented by a triple $G =(V,E,\sigma)$, where $V$ is the set of the $n$ nodes, $E$ is the set of the undirected edges $e=(v,w)$ with $v,w\in V$, and $\sigma$ is a mapping $\sigma: E\to \{\pm 1\}$ which assigns to each edge a sign. The signed network can be completely characterized by the $n \times n$ signed adjacency matrix ${A}=[A_{vw}]$.  Specifically, if there is no edge $(v,w)$, $A_{vw}=0$; otherwise, $A_{vw}=+1$ represents a positive edge, and $A_{vw}=-1$ represents a negative edge.
We assume graphs without self-loops so that the matrix ${A}$ has null diagonal entries.
The underlying unsigned network, i.e. the graph obtained from $G$ by ignoring the sign of the edges, is associated with the matrix $|{A}|$ of the absolute value entries of ${A}$.
The absolute degree of a node $v$, denoted by $k_{v}$, is defined
to be the number of edges incident to that node, irrespective of the edge signs, that is $k_{v}=\sum_{w} |A_{vw}|$.
A non-empty trail in which only the first and the last node coincide is called cycle. The sign of a cycle is the product of the signs of its edges.
A signed graph $G$ is \textit{structurally balanced} if there are no negative cycles, or, equivalently, there are no cycles with an odd number of negative edges (see \cite{cartwright1956}). In a balanced network, the set $V$ can be partitioned into two subsets such that every edge between them is negative, while every edge within them is positive (see \cite{harary1953}).

\section{Part I. Local balance index}

\subsection{Motivation of the local balance index}

\begin{figure}
	\begin{centering}
		\includegraphics[width=0.6\textwidth]{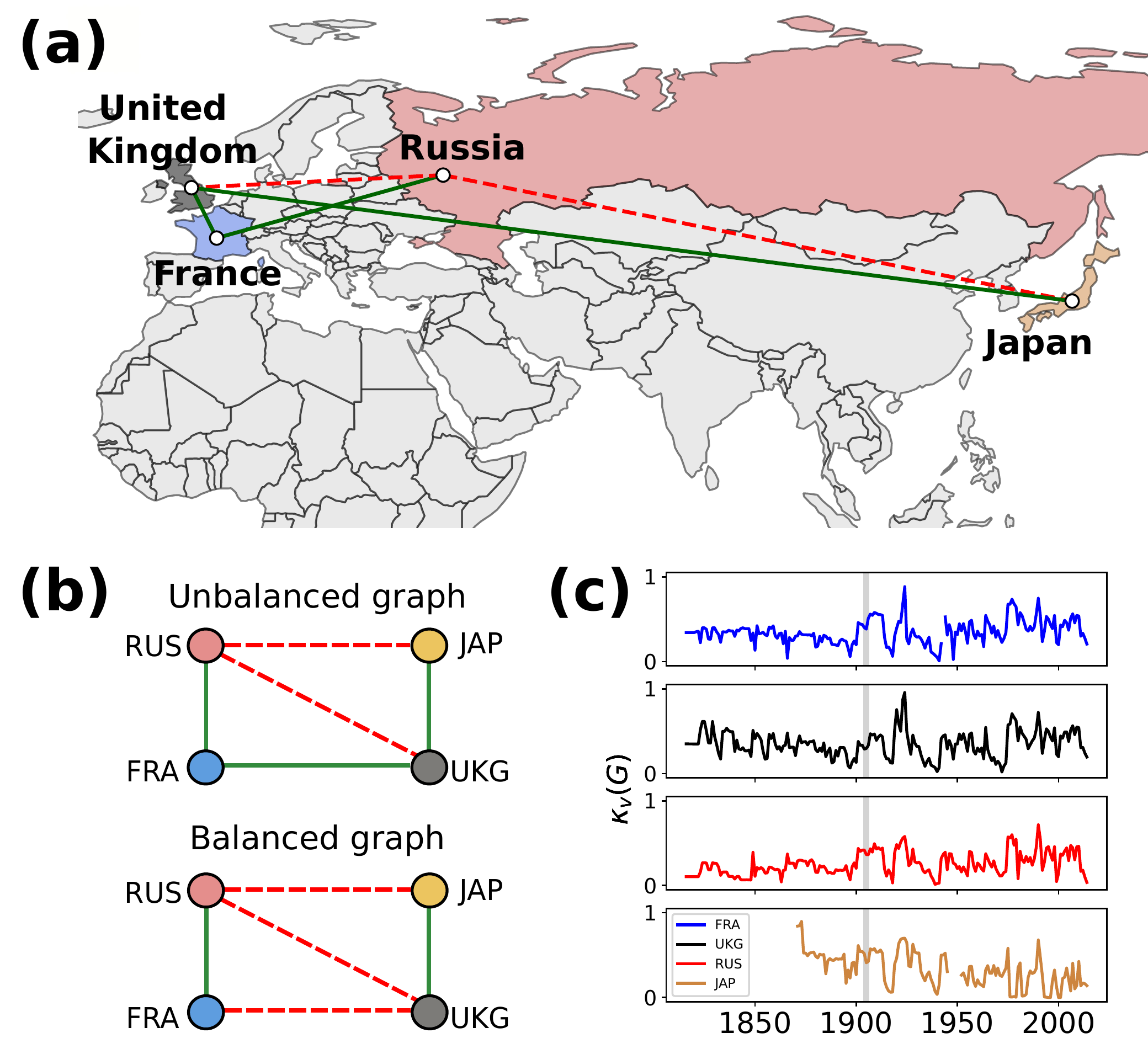}
		\par\end{centering}
	\caption{(a) Illustration of the relations between Great Britain, France, Russia
		and Japan during the Russo-Japanese war (1904-05). (b) The mathematical
		framework for the theory is the representation of IR as signed networks,
		in which positive edges (green solid lines) represent interstates
		alliances and negative ones (red dashed lines) are reserved for any
		kind of conflict/enmities between the parts. (c) Temporal evolution
		of the \textquotedblleft local balance\textquotedblright{} of the countries
		in the examined time period. The gray area highlights the years in
		which the Russo-Japanese war took place.}
	
	\label{First_Example}
\end{figure}

Let us motivate the necessity of defining a local balance index with
one example from the real-world. In the period 1904-1906 France and
Russia were allies (Franco-Russian treaty), as were Japan and Britain (Anglo-Japanese alliance) and France and Britain (Entente Cordiale). However, Russia and Britain were
enemies at the time when Russia went to war with Japan. Therefore,
France, Great Britain and Russia formed a signed triangle \cite{Signed_graphs}
with two positive and one negative edge (see Fig. \ref{First_Example}
(a)). Then, if Britain made effective its support to Japan, Russia
may declare war against its enemy Great Britain. This would oblige
France to make effective its support to Russia as they were allies,
but to enter into a contradiction with its new ally, the United Kingdom. This
conflictive situation is a consequence of the fact that the triangle
UK-Russia-France is unbalanced (Fig. \ref{First_Example} (b), top
panel). The real consequence was that neither
France supported Russia during the war, nor Great Britain made
its alliance with Japan effective. These contradictions would not exist if the
triangle between these countries were balanced (Fig. \ref{First_Example}
(b), bottom panel), where two factions (Russia-France vs. Japan-UK)
would be formed. 

As can be seen in Fig. \ref{First_Example} (b), top panel, UK-Japan-Russia
formed a balanced triangle. Consequently, in total, Japan participates in one balanced
triangle and in zero unbalanced ones, Russia and Great Britain are part of one
balanced and one unbalanced triangle, and France participates in an unbalanced
one. Thus, not all countries have the same balancing position in this
historic event. To capture these differences in the contribution of
individual nodes to the global balance of the network we will define
a local balance index.

\subsection{Definition and mathematical properties of the local balance index}

A closer look at Fig. \ref{First_Example} (b), top panel, reveals
that France is not only related to the rest of conflicting nations
via a balanced triangle, but also by means of the unbalanced square
RUS-JAP-UKG-FRA. The importance of considering cycles beyond triangles
for characterizing the degree of balance of a network is well known
since the pioneering works of Cartwright and Harary \cite{cartwright1956}
(see also \cite{Estrada_2019}). Instead of
counting signed cycles, which is a computationally costly procedure,
we will quantify the local balance by counting local closed walks
(LCW). A LCW of length $k$ starting at node $v$ in a signed network
$G$ is given by $\left(A^{k}\right)_{vv}$, where $A$ is the adjacency matrix of the network. The diagonal elements of the $k-$power of this matrix can also be written as: $(A^{k})_{vv}=\textnormal{\# of positive LCW of length }\ensuremath{k}\ensuremath{-}\ensuremath{\lvert\textnormal{\# of negative LCW}\textnormal{ of length }\ensuremath{k}\rvert}.$ 

Let us consider a LCW of length $k$ with one negative edge $e=\left\{ p,q\right\} $.
Every node $v\neq p,q$ is at distance $i$ from $p$ and at distance
$n-i-1$ ($i=1,\ldots,n-2$) from $q$. This means that the longer
the LCW, the smaller the average influence of the negative edge on
the rest of the nodes of the cycle. Among the several choices giving rise to specific matrix functions \cite{higham2008functions}, we choose here to penalize every LCW
of length $k$ by $1/k!$, such that we have the following definition.
\begin{definition}
	Let $G$ be any signed graph with adjacency matrix $A$.
	Let $\lvert A\rvert$ be the entrywise absolute of the adjacency matrix
	$A$ of the network. The local balance index of $v\in V$ is defined
	as:
\end{definition}

\begin{align}
	\kappa_{v}(G):=\frac{\sum_{k=0}^{\infty}(A^{k})_{vv}/k!}{\sum_{k=0}^{\infty}(\lvert A\rvert^{k})_{vv}/k!}=\dfrac{(\exp A)_{vv}}{(\exp\lvert A\rvert)_{vv}},\label{local_balance_index}
\end{align}

We now prove some general properties of the local balance index.
\begin{proposition}
Let $G$ be any signed connected graph and $v\in V$.
Then, $\kappa_{v}(G)$ is bounded as follows:
	
	\begin{equation}
		0<\kappa_{v}(G)\leq1.
	\end{equation}
\end{proposition}

\begin{proof}
	Let $\alpha_{1}>\alpha_{2}>\ldots >\alpha_{n}$ and $a_{1}, a_{2},\ldots ,a_{n}$ be the eigenvalues
	and associated eigenvectors of $A$, respectively. Similarly, let
	$\beta_{1}>\beta_{2}>\ldots >\beta_{n}$ and $b_{1},b_{2},\ldots,b_{n}$ be the eigenvalues
	and eigenvectors of $|A|$, respectively. Then, we can write Eq. \eqref{local_balance_index}
	as: 
	\begin{align}
		\kappa_{v}(G)=\frac{\sum_{j=1}^{n}(a_{j})_{v}^{2}e^{\alpha_{j}}}{\sum_{l=1}^{n}(b_{l})_{v}^{2}e^{\beta_{l}}},\label{spectral_dec}
	\end{align}
	where $(a_{j})_{v}^{2}$ denotes the square of the $v$th entry of
	the vector $a_{j}$. From eq. \eqref{spectral_dec}, it is clear that
	$\kappa_{v}(G)>0$ for any $v$, since all terms within each sum are
	nonnegative and at least one of the $(a_{j})_{v}$ is different from
	zero. \\
	To prove that $\kappa_{v}(G)\leq1$, we use the power-series expansion
	of the exponential function: $(e^{A})_{vv}=\sum_{k}\frac{(A^{k})_{vv}}{k!}$.
	We recall that $(A^{k})_{vv}$ counts the difference between the number
	of positive and negative closed walks starting and ending at node
	$v$. Let us denote by $\left(W^{+}\right){}_{vv}^{k}$ and $\left(W^{-}\right){}_{vv}^{k}$
	the number of positive and negative walks of length $k$ starting
	and ending at the node $v$, respectively. Then, the local balance
	index can be rewritten as: 
	\begin{align}
		\kappa_{v}(G)=\frac{\sum_{k=0}^{\infty}\frac{1}{k!}[(W^{+})_{vv}^{k}-(W^{-})_{vv}^{k}]}{\sum_{k=0}^{\infty}\frac{1}{k!}[(W^{+})_{vv}^{k}+(W^{-})_{vv}^{k}]}.\label{walks}
	\end{align}
	Since $(W^{+})_{vv}^{k}-(W^{-})_{vv}^{k}\leq (W^{+})_{vv}^{k}+(W^{-})_{vv}^{k}$, then $\kappa_{v}(G)\leq1$. 
\end{proof}
Let us now recall two basic concepts of balance theory on signed graphs. 
\begin{definition}
	Let $G$ be any signed, connected graph and let $C$
	be any cycle in $G.$ The sign of the cycle $C$ is the product of
	the signs of all edges in the cycle. $C$ is called positive if its
	sign is positive, otherwise it is negative. 
\end{definition}

\begin{definition}
	Let $G$ be any signed, connected graph. The signed
	graph $G$ is balanced if and only if, all of its cycles are positive. 
	
\end{definition}
We then have the following property of the local balance index.

\begin{proposition}
	Let $G$ be any signed connected graph. $G$ is balanced
	if and only if $\kappa_{v}(G)=1$ for any node $v$. 
\end{proposition}

\begin{proof}
	Recall that the global balance index of $G$ is defined as \cite{Estrada_2014}
	
	\begin{equation*}
		\kappa(G):=\dfrac{tr\left(e^{A}\right)}{tr\left(e^{\left|A\right|}\right)},
	\end{equation*}
	where $tr$ is the trace of the corresponding matrix. Then, we can
	write: 
	\begin{align}
		\kappa\left(G\right)=\frac{\sum_{v=1}^{n}\left(\kappa_{v}\left(G\right)SC_{v}\right)}{\sum_{v=1}^{n}SC_{v}},\label{relation_local_global}
	\end{align}
	where $SC_{v}=(e^{|A|})_{vv}$ is the subgraph centrality of a node
	$v$ \cite{Estrada_2005}. It has been proved in previous works that $\kappa(G)=1$
	if and only if $G$ is balanced (see \cite{Estrada_2014}). By Eq. \eqref{relation_local_global}, this is true if and only if and only if $\kappa_{v}\left(G\right)=1$ for all $v\in V$. 
\end{proof}
The lower bound of $\kappa_{v}(G)$ can be improved by virtue
of the following proposition. 
\begin{proposition}
	Let $G$ be any signed connected graph and let $k_{max}=\underset{v\in V}{\max}\, k_{v}$.
	Then,
	
	\begin{equation}
		\kappa_{v}(G)\geq e^{-2k_{max}}.
	\end{equation}
\end{proposition}

\begin{proof}
	Let us denote the smallest eigenvalue of $A$ by $\alpha_{min}$.
	Then, using the spectral decomposition of $A$, we obtain the following
	inequality: 
	\begin{align*}
		(e^{A})_{vv}=\sum_{j=1}^{n}(a_{j})_{v}^{2}e^{\alpha_{j}}\geq e^{\alpha_{min}}\sum_{j=1}^{n}(a_{j})_{v}^{2}=e^{\alpha_{min}}
	\end{align*}
	In the last step, we have used the fact that $\sum_{j}(a_{j})_{v}^{2}=1$,
	which follows from the orthonormality of the eigenbasis $\{a_{j}\}$.
	To find a lower bound for $\alpha_{min}$, we use Gershgorin's theorem.
	According to it, all eigenvalues of $A$ lie on the union of the disks
	$D_{v}$ centered at $A_{vv}$ and with radius $R_{v}=\sum_{w\neq v}|A_{vw}|$.
	Because $A$ is the adjacency matrix of a graph without self-loops,
	$A_{vv}=0$ for all $v$ and $R_{v}=\sum_{w\neq v}|A_{vw}|=k_{v}$,
	i.e., the radius of each Gershgorin disk is the absolute degree of
	the corresponding node. Consequently, all eigenvalues must lie within
	a disk centred at the origin and with radius $k_{max}$. As a consequence,
	$\alpha_{min}\geq-k_{max}$ and $(e^{A})_{vv}\geq e^{-k_{max}}$.
	Repeating the same calculation for $e^{|A|}$, we find that $(e^{|A|})_{vv}\leq e^{\beta_{max}}\leq e^{k_{max}}$.
	Finally, combining these two inequalities results in $\kappa_{v}(G)=\frac{(e^{A})_{vv}}{(e^{|A|})_{vv}}\geq e^{-2k_{max}}$. 
\end{proof}
\begin{corollary}
Let $G$ be any signed connected graph.  Then, $\kappa_{v}(G)\geq e^{2-2n}$. 
\end{corollary}

We now find which graph has the lowest possible $\kappa_{v}(G)$.

\begin{proposition}
	Let $K_{n}^{-}$ be the complete graph with all-negative edges. Then,
	\begin{equation}
		\lim_{n\to\infty}\kappa_{v}(K_{n}^{-})=0, \ \forall v \in V.
	\end{equation}
\end{proposition}

\begin{proof}
	The spectrum of the complete graph is $Sp(K_{n})=\{(-1)^{[n-1]},(n-1)^{[1]}\}$,
	while the spectrum of the complete all-negative graph is $Sp(K_{n}^{-})=\{(1-n)^{[1]},1^{[n-1]}\}$
	(the superindices indicate the multiplicity of each eigenvalue). Since
	$K_{n}^{-}$ is a complete all-negative graph, all nodes are equivalent
	and therefore $(e^{A})_{vv}$ is independent of $v$. Consequently,
	$(e^{A})_{vv}=\frac{tr(e^{A})}{n}=\frac{e^{1-n}+(n-1)e}{n}$. Similarly,
	$(e^{|A|})_{vv}=\frac{e^{n-1}+(n-1)e^{-1}}{n}$. Substituting in the
	expression of the local balance index defined in Eq. \eqref{local_balance_index}, we finally get: 
	\begin{equation*}
		\kappa_{v}(K_{n}^{-})=\frac{{e^{1-n}}+(n-1)e}{{e^{n-1}}+(n-1)e^{-1}},
	\end{equation*}
	which becomes zero in the limit $n\to\infty$. 
\end{proof}

The following result regarding the local balanced of cycle graphs \cite{akbari2018spectral} is proved in Sec. 1 of the SM.

\begin{proposition}
	\label{signedcycles}
	Let $C_{n}^{k-}$ be a signed cycle of length $n$ and $k$ negative
	edges. Then, the local balance index of any node in $C_{n}^{k-}$
	is
	
	\begin{equation}
		\kappa_{v}\left(C_{2l}^{k-}\right)= \begin{cases}
			\dfrac{\sum_{j=0}^{2l-1}\exp\left({2\cos\left(\dfrac{\left(2j+1\right)\pi}{2l}\right)}\right)}{\sum_{j=0}^{2l-1}\exp\left({2\cos\left(\dfrac{j\pi}{l}\right)}\right)} & k\textnormal{ odd},\\
			1 & k\textnormal{ even},
		\end{cases}
	\end{equation}
	
	for even cycles and
	
	\begin{equation}
		\kappa_{v}\left(C_{2l+1}^{k-}\right)= \begin{cases}
			\dfrac{\sum_{j=0}^{2l}\exp\left({2\cos\left(\dfrac{2j\pi}{2l+1}\right)}\right)}{\sum_{j=0}^{2l}\exp\left({2\cos\left(\dfrac{\left(2j+1\right)\pi}{2l+1}\right)}\right)} & k\textnormal{ odd},\\
			1 & k\textnormal{ even},
		\end{cases}
	\end{equation}
	
	for odd ones.
\end{proposition}

\begin{remark}
	The previous result illuminates several important structural aspects
	of the local balance in cycles. First, that every node in a signed
	cycle will have the same local balance independently of how close
	it is from the negative edges. Second, that the parity of the number
	of negative edges but not their number is what determines balance
	in signed cycles. Third, that for sufficiently large $n$ the presence of the negative edge becomes negligible, fulfilling our intuition that the umbalance is ``diluted'' in very large cycles. This statement is expressed in the following corollary.
\end{remark}

\begin{corollary}
Let $C_{n}^{k-}$ be a signed cycle of length $n$ and $k$ negative edges. Then
$\kappa_v\left(C_{n\rightarrow\infty}^{k-}\right)\rightarrow1$.
\end{corollary}

Finally, we give a result, also proved in Sec. 1 of the SM, concerning a type of graph whose structure is found in several practical applications, such as NIR, as we will soon show.

\begin{definition}
	Let $K_{n}$ be a complete graph with $n$ nodes. Let $l\leq n-1$.
	Then, pick $l$ nodes in $K_{n}$ to form the set $S$ and change
	the sign of their edge $(v_i,v_j)$ to negative if and only if $\left\{ v_{i},v_{j}\right\} \subset S$.
	Then, the set $S$ forms a negative clique of $l$ nodes in $K_{n}$ and we will denote the resulting graph as $K_{n}(K_{l}^{-})$.
\end{definition}

\begin{proposition}
	\label{negativeclique}
	Let $K_{n}(K_{l}^{-})$ be a graph defined as before.  $\kappa_{v}\left(K_{n}(K_{l}^{-})\right)$
	tends to zero when $l\to\infty$ and $m:=n-l$ remains finite. 
\end{proposition}

\subsection{Local balance and diffusion of 'information'}
We now show how the local balance index can be obtained in an alternative way as a result of a non-conservative diffusion process. 
It has been claimed that diffusion \cite{masuda2017random} of information, generally speaking,
in social systems can be a non-conservative process \cite{NC_diffusion_2,NC_diffusion_3,NC_diffusion_5}.
By non-conservative it is meant that the amount of information at
a given time can be smaller/bigger than that at the initial time.
Lerman and Ghosh \cite{Lerman-Ghosh} defined the non-conservative
Laplacian of a network as a way to model such kind of diffusive processes.
In the case of a signed network with adjacency matrix $A$ the Lerman-Ghosh
Laplacian could be read as: $\mathcal{L}:=\chi I-A$ and for the unsigned
version of the same network, we should define $\tilde{\mathcal{L}}:=\chi I-\left|A\right|$,
where $\chi$ is a scalar parameter and $I$ is the identity
matrix.

Then, the non-conservative diffusion on the signed graph is described
by

\begin{equation}
	\dot{u}\left(t\right)=-\mathcal{L}u\left(t\right),u\left(0\right)=u_{0},
	\label{dynamics}
\end{equation}
where $u(t)$ is the state vector at time $t$ and $u_{0}$ is the initial state. The solution of Eq. \eqref{dynamics} is then
\begin{equation*}
	u\left(t\right)=e^{-\mathcal{L}t}u_{0}=e^{-\chi t}e^{tA}u_{0}.
\end{equation*}
This solution can be split into two terms:
\begin{equation*}
	e^{-\chi t}\left\{ \left[\begin{array}{c}
		\left(e^{tA}\right)_{11}(u_{0})_{1}\\
		\vdots\\
		\left(e^{tA}\right)_{nn}(u_{0})_{n}
	\end{array}\right]+\left[\begin{array}{c}
		\left(e^{tA}\right)_{12}(u_{0})_{2}+\cdots+\left(e^{tA}\right)_{1n}(u_{0})_{n}\\
		\vdots\\
		\left(e^{tA}\right)_{n1}(u_{0})_{1}+\cdots+\left(e^{tA}\right)_{nn-1}(u_{0})_{n-1}
	\end{array}\right]\right\} ,
\end{equation*}
where the first term accounts for the amount of diffusive information
that departs from a node and returns to (or is retained at) it.

Obviously, we can write down the same equation for the unsigned version
of the network, such that:

\begin{equation*}
	\dot{\tilde{u}}\left(t\right)=-\tilde{\mathcal{L}}\tilde{u}\left(t\right),\tilde{u}\left(0\right)=u_{0},
\end{equation*}
with the same initial condition as before. Thus,
$\tilde{u}\left(t\right)=e^{-\chi t}e^{t\left|A\right|}u_{0}.$

If we want to know how much information is "lost" in a non-conservative
diffusion on a given network as a consequence of the existence of
signed edges in it, we can calculate the ratio $u_{v}\left(t\right)/\tilde{u}_{v}\left(t\right)$
with initial condition $(u_{0})_{i}=\delta_{iv}$ (where $\delta_{iv}$ denotes the Kronecker delta),
such that

\begin{equation*}
	\frac{u_{v}\left(t\right)}{\tilde{u}_{v}\left(t\right)}=\dfrac{e^{-\chi t}\left(e^{tA}\right)_{vv}}{e^{-\chi t}\left(e^{t\left|A\right|}\right)_{vv}}=\dfrac{\left(e^{tA}\right)_{vv}}{\left(e^{t\left|A\right|}\right)_{vv}}.
\end{equation*}

Therefore, the local balance index can be interpreted as the ratio of the information
``lost" in a non-conservative diffusion on a signed network when $t=1$.

Finally, we remark that the parameter $t\in{\mathbb R}^{+}$ can also be interpreted as an overall weight that is applied to every edge of the network.
Consequently, we can define a weighted version of the local balance
index:
\begin{equation*}
	\kappa_{v}\left(G,t\right):=\dfrac{\left(e^{tA}\right)_{vv}}{\left(e^{t|A|}\right)_{vv}},
\end{equation*}
and we have the following result:
\begin{proposition}
	\label{weighted_proposition} Let $\kappa_{v}\left(G,t\right)$ be the local
	balance index of a node $v$ in an unbalanced graph $G$, where every
	edge is weighted by a factor $t\in{\mathbb R}^{+}$. Then,
	
	\begin{equation}
		\lim_{t\rightarrow\infty}\kappa_{v}\left(G,t\right)=0,
	\end{equation}
	\begin{equation}
		\lim_{t\rightarrow0}\kappa_{v}\left(G,t\right)=1.
	\end{equation}
\end{proposition}

\begin{proof}
	On the one hand, 
	\begin{align*}
		& \lim_{t\rightarrow\infty}\kappa_{v}\left(G,t\right)=\lim_{t\rightarrow\infty}\dfrac{(a_{1})_{v}^{2}e^{t\alpha_{1}}}{(b_{1})_{v}^{2}e^{t\beta_{1}}}=\lim_{t\rightarrow\infty}\dfrac{(a_{1})_{v}^{2}}{(b_{1})_{v}^{2}}e^{t\left(\alpha_{1}-\beta_{1}\right)}=0,
	\end{align*}
	because $\beta_{1}>\alpha_{1}$ (theorem 3.9 in \cite{tian2022spreading}).
	On the other hand, 
	\begin{align*}
		& \lim_{t\rightarrow0}\kappa_{v}\left(G,t\right)=\dfrac{\sum_{j=1}^{n}(a_{j})_{v}^{2}}{\sum_{j=1}^{n}(b_{j})_{v}^{2}}=1.
	\end{align*}
\end{proof}

\section{Part II. Application of local balance to global international relations}

\begin{figure}[h]
	\includegraphics[width=1\textwidth]{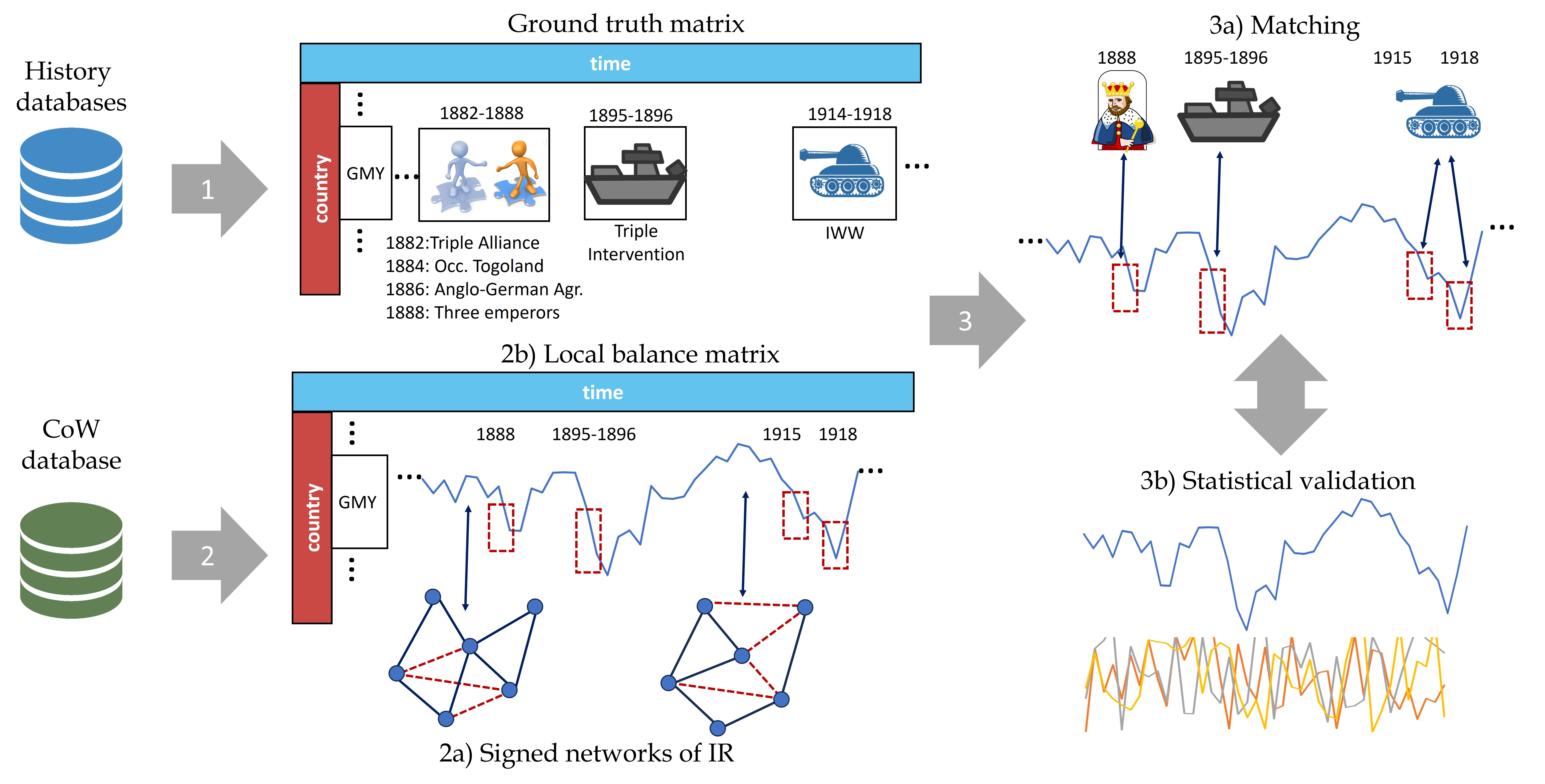}
	
	\caption{Illustration of the general scheme for relating the local balance
		index of signed networks with major historic events which have occurred
		in the World in the period 1816-2014. First, a ground truth matrix
		of events is created from databases of historic events for each of
		the 217 countries analyzed (1). Independently, the networks of international
		relations (2a) are created from the information provided at the CoW
		database and the time series of country's local balance are built
		(2b). Meaningful peaks and valleys in the time series of local balance
		are detected (2b). Then, a matching process (3a) allows to pair the
		peaks/valleys detected in the time series with the events recorded
		in the ground truth. The statistical significance (3b) of these matchings
		is then analyzed and reported.}
	
	\label{general scheme} 
\end{figure}

The main goal of this second part of the paper is to apply the local
balance index to an in-depth analysis of the global international relations
(IR) for the period between 1816 and 2014. While previous analysis
of the IR for similar periods have focused on the study of the global
balance of signed networks \cite{Estrada_2019, kirkley2019balance}, here we concentrate our analysis on the
countries' role on this global balance. Therefore, by an in-depth
analysis we refer to the fact that we will study the connection between
the countries' local balance index--a purely topological index derived
from signed networks of IR--with major historical events involving
these countries in specific periods of time. For conducting this analysis
we base the current study in the general methodological scheme illustrated
in Fig. \ref{general scheme}. We detail the methodological aspects
of each step of this scheme in the following separate subsections.

\subsection{Methodology and statistical assessment}

\subsubsection{Construction of signed networks of IR}

For every year in the period 1816-2014 we construct a signed network
of IR, which corresponds to the step 2a in Fig. \ref{general scheme}.
The nodes of each network correspond to a subset of the 217 countries
studied (see Table \ref{codes2} in the SM for a list). The connections between the countries correspond to signed
relations reported in databases different from the sources used for
constructing the ground truth matrix. These databases were collected
from \cite{Methods_1,Methods_2} for alliances, \cite{Methods_3}
for enmities, and \cite{Methods_4} for strategic rivalries. The coding
of inter-country relations is as follows. Two states are assigned
a score of $+1$ at year $t$ if they had any type of alliance (defense
pact, offense pact, nonaggression pact, neutrality pact) during that
year and zero otherwise. Consultation pacts (see \cite{Methods_2})
are excluded. On the other hand, a dyad
is assigned a score of $-1$ at time $t$ if members had militarized
interstate disputes at time $t$ or were considered to be strategic
rivals at this time according to the Colaresi et al. \cite{Methods_4}
data. Note that alliances may be asymmetric (see \cite{Methods_1} for further explanation), while all enmities are symmetric. Since the number of asymmetric
edges was very small, we converted data for each year into an undirected,
unweighted, signed network where nodes represent countries and edges
their relations in a given year. If two countries happen to be part
of both an alliance and a conflict in the same year, we consider the
conflict to be dominant, and therefore the edge to be negative.

Data are thus arranged in a temporal network spanning 199 years with
a total of 217 nodes. We emphasize that the number of nodes is far
from constant over time, as new states are created and old empires
disappear. In fact, the number of nodes typically increases, from
a minimum of 23 nodes in 1816 to a maximum of $195$ nodes in 2014.
The average number of nodes per year is 66. The edge statistics follows
a similar trend: it increases from 88 links in 1816 to 8,124 in 2014,
with an average number of 1,213 edges per year. On average, $41.22\%$
of the edges are negative, although this number also fluctuates over
time: a minimum of $1.49\%$ negative interactions is reached in 2007,
while the maximum of $97.30\%$ is observed in 1880. It should also
be noted that annual networks are typically not connected and often
have many isolated nodes. The giant component represents on average
$69.42\%$ of the entire network, with a rather exceptional minimum
of $33.33\%$ in 1871 and a maximum of $95.38\%$ in 2011 and 2013.
Over the last 50 years, the giant component always exceeds $90\%$
of the whole network. Since it selects the main actors on the geopolitical
scene, the individual local balance values have been computed for
such a component only.

\subsubsection{Construction of ground truth matrix}

First we create an empty matrix $\Gamma$ with $N_{c}=217$ rows, each one corresponding
to one of the 217 countries analyzed in this work. Every one of the
$N_{t}=199$ columns corresponds to one of the years in the period
1816-2014 which are covered by this study. Then, we populate this
matrix by using information from different sources 
about historic events involving a given country for specific years
(see step 1 in Fig. \ref{general scheme}). Notice that these sources of historic data are independent of those used before to create the NIR. Historical events are
classified into two groups: 'peaceful' events and 'conflicting' events.
Peaceful events are assigned a (+1) sign while conflicting events are assigned a (-1) sign. Assigning (+1) to peaceful events and (-1) to conflicting events is a conventional agreement. Reverting such assignment
would not affect the obtained results.
Peaceful events include: 
\begin{enumerate}[label=\roman*.]
	\item The establishment of treaties, accords or agreements after wars; 
	\item The implementation of democratic procedures, like elections, change
	of regime or adoption of constitutions; 
	\item Countries becoming independent from their colonial domination. 
\end{enumerate}
On the other hand, conflicting events include: 
\begin{enumerate}[label=\roman*.]
	\item The outbreak of an armed or trade conflict between two or more countries,
	or the entry of a country into an existing conflict, and in some cases
	clearly identified battles, invasions, or expeditions; 
	\item Revolutions, insurrections, or civil wars within a single nation,
	but with outside nations involved in support of certain warring factions.
	In this context, wars of independence from foreign domination, such
	as those in 19th-century Europe or 20th-century Africa, deserve special
	mention; 
	\item The effects of coups d'etat, which alter geopolitical arrangements
	and relations with foreign nations; 
	\item Crises in the broad sense, that is, the institutional, military, or
	economic collapse of a given nation brought about by international
	political conditions. 
\end{enumerate}
For example, given that Germany (GMY) is one of the participants in
the Triple Alliance of 1882, we assign a (+1) sign to the entry corresponding
to the pair (GMY, 1882) in the ground truth matrix. Similarly, due
to the French involvement in the Sino-French War (1884-1885), we assign
a (-1) sign to the pair (FRN, 1884). 

To be consistent with the IR database (see next section), we only
include events if the involved country appears in the IR database
in the year of the event and in the previous one. For example, the
Indian Rebellion of 1857 does not translate to a (-1) in (IND, 1857),
because India is not considered an autonomous entity in the IR dataset
until 1947 (notice, however, that this event does translate into a
(-1) in the entry (UKG, 1857)). 

In total, we have considered 1533 major historic events. The countries
participating in the largest number of historic events are: UK (76), Russia (61), USA
(61), France (52), Italy (45), Germany (44), Turkey (43) and China
(42). They coincide with the major actors of history during the centuries
XIX-XXI. There are 47 countries, mainly small ones, for which there
are no recorded major historical event. The years with the largest
number of events are: 1999 (44), 1992 (34), 1979 (34), 1989 (33),
1975 (28), 1991 (27), 1990 (26), 2011 (24), and 1967 (23). There are
only 11 years for which there is no major historic event recorded,
all of them in the XIX century.

\subsubsection{Detection of peaks and valleys}

Here we construct a local balance matrix $T$ of dimensions $N_{c}\times N_{t}$ based on the detection of peaks and valleys, that is positive and negative changes, in the temporal series of the local balance index of every country in its signed network
(see step 2b in Fig. \ref{general scheme}). That is, for each year
$t$, we extracted the signed adjacency matrix of the network and
calculated the local balance $\kappa_{v}(t)$ for each country $v$.
Afterwards, we aggregated the data from every year, obtaining a time
series of the local balance values for each country (see, for instance, Fig. \ref{First_Example}, panel (c)). Then, we computed
the difference in local balance for consecutive years, $\Delta\kappa_{v}(t)=\kappa_{v}(t)-\kappa_{v}(t-1)$,
and identified the largest decreases or increases. We assign a (-1)
sign to a country-year pair if $\Delta\kappa_{v}(t)\leq-0.1$ and
$\kappa_{v}(t)\leq+0.5$ , and a (+1) sign if $\Delta\kappa_{v}(t)\geq0.1$
and $\kappa_{v}(t)\geq+0.5$. The conditions included for $\kappa_{v}(t)$,
e.g. $\kappa_{v}(t)\geq+0.5$ and $\kappa_{v}(t)\leq+0.5$ respectively,
guarantee that the system at the analyzed time has a significant degree
of (un)balance. Since the method relies on differences in $\kappa_{v}$,
events occurring in the year in which a country enters the dataset
cannot be detected, as the balance for the year $t-1$ cannot be calculated.

We identified 1829 valleys and 834 peaks. However, the events are
not homogeneously distributed over the time range. For instance, the
countries with the largest number of valleys across the period studied
are: Iran (34), Turkey (34), UK (32), USA (30), France (29), Yugoslavia
(28), Italy (27), Greece (27), Japan (27), Spain (26), Romania (24)
and Germany (24). There are 17 countries without any valley. On the
other hand, the countries with the largest number of peaks are: USA
(27), Italy (23), Romania (21), Greece (20), UK (19), Portugal (19)
and Spain (18). There are 52 countries for which no peaks were detected
based on the criterion used here.

\subsubsection{Analysis of the matching of peaks/valleys with historic events}

We are now in the position to test the similarities between the ground
truth matrix $\Gamma$ and the local balance matrix $T$. We employ two
similarity criteria: the Pearson correlation coefficient between matrices $\Gamma$ and $T$:
\begin{align}
	s_P(\Gamma,T) = \frac{Cov(\tilde \Gamma, \tilde T)}{\sqrt{Var(\tilde \Gamma)Var(\tilde T)}}
\end{align}
(where $\tilde \Gamma$ and $\tilde T$ are one-dimensional vectors obtained by concatenating the rows  of $\Gamma$ and $T$ in a standard matrix flattening procedure),
and the Frobenius similarity of the pairs of matrices:
\begin{align}
	s_F(\Gamma,T)= \frac{tr(\Gamma^TT)}{\sqrt{tr(\Gamma^T\Gamma)tr(T^TT)}}  . \label{eq:similarity1}
\end{align}
The Frobenius similarity is the
equivalent for matrices to the cosine similarity index between two
vectors. In Sec. 2 of the SM, we give a more detailed description of these two similarity indices.

We need now to check whether the similarities found between the ground
truth and local balance matrices are statistically significant. For
this test we create the null model matrix $T'$ as follows. For each
country we randomize the corresponding time series 500 times, and
then we construct the matrix $T'$ by applying the methodology that
was previously described for the matrix $T$. Given that through the
time range analyzed, many countries have been created or have disappeared,
the balance time series for many countries contains missing data,
corresponding to the years in which those countries did not exist.
Consequently, we exclude the missing values from the shuffling; that
is, we keep the periods of missing data fixed and only randomize the
time periods where the given country exists.

Using $s_P$ and $s_F$ as similarity
indices, we find that the matrix of peaks and valleys and the ground truth matrix have a statistically significant similarity, ten times bigger than that of the randomized truth model (0.306 vs. 0.039 for Pearson and 0.313 vs. 0.052 for Frobenius). Both quantitative similarity measures indicate
that the matrix of peaks and valleys based on the real time series
of local balance is significantly more similar to the ground truth
of historic events than its randomization. To test whether these similarity
indices are significantly different or not we use the statistical
tests included in \cite{diedenhofen2015cocor}. All of them confirmed that the similarity with the real matrix is significantly larger than that with the randomized one. We have
also tested the similarity between the absolute values of the matrices;
i.e., indicating only the existence of a major event and not its nature,
and we also obtain that the correlations with the real local balance
matrix are statistically more significant than the correlations with
the randomized matrix.

\subsubsection{Correlations with GeoPolitical Risk indices}

As a further test of the statistical significance of the method developed
here, we check for several individual countries the connection between
the local balance and the perceived geopolitical risk. This analysis
is based on the GeoPolitical Risk (GPR) index \cite{CaldaraIacoviello2022},
which measures the frequency of newspaper articles that discuss adverse
geopolitical events. We employ the historic variant, that aggregates
the data of three different U.S. newspapers to provide monthly measures
of the risk of 44 different countries between the years 1900 and 2023.
Then, we calculate the Spearman's correlation coefficient \cite{Spearman1904}
between GPR and local balance time series for the nine countries whose
time overlap includes 100 or more years: China, France, United Kingdom,
Italy, Japan, Portugal, Russia/USSR, Turkey and the USA. We also compare
the correlation between the GPR time series and 500 randomizations
of each local balance time series, following the procedure outlined
in Sec. 4.1.4. The resulting rank correlation coefficients are
provided in Table \ref{GPR} in the SM, where we observe significant correlations for most of the countries. For instance, Spearman rank correlation
for the USA, France and Portugal are -0.548, -0.414 and -0.367, respectively,
while for the same countries in the analogous random networks the
rank correlations are 0.000486, -0.001147 and 0.006732, respectively.
Interestingly, the weakest correlations are found in non-European
countries, a fact that suggests a possible Western bias in the U.S.
newspapers used to compute the GPR index.

Based on the similarity criteria between the ground truth matrix and
the matrix of peaks and valleys, and the confidence provided by the
correlation of the local balance index with GPR we proceed to match
the peaks and valleys with the corresponding historic events that
match them in terms of countries and exact year of occurrence.

\subsection{Linking local balance with historical narrative}

\begin{figure}[h]
	\begin{centering}
		\includegraphics[width=0.9\textwidth]{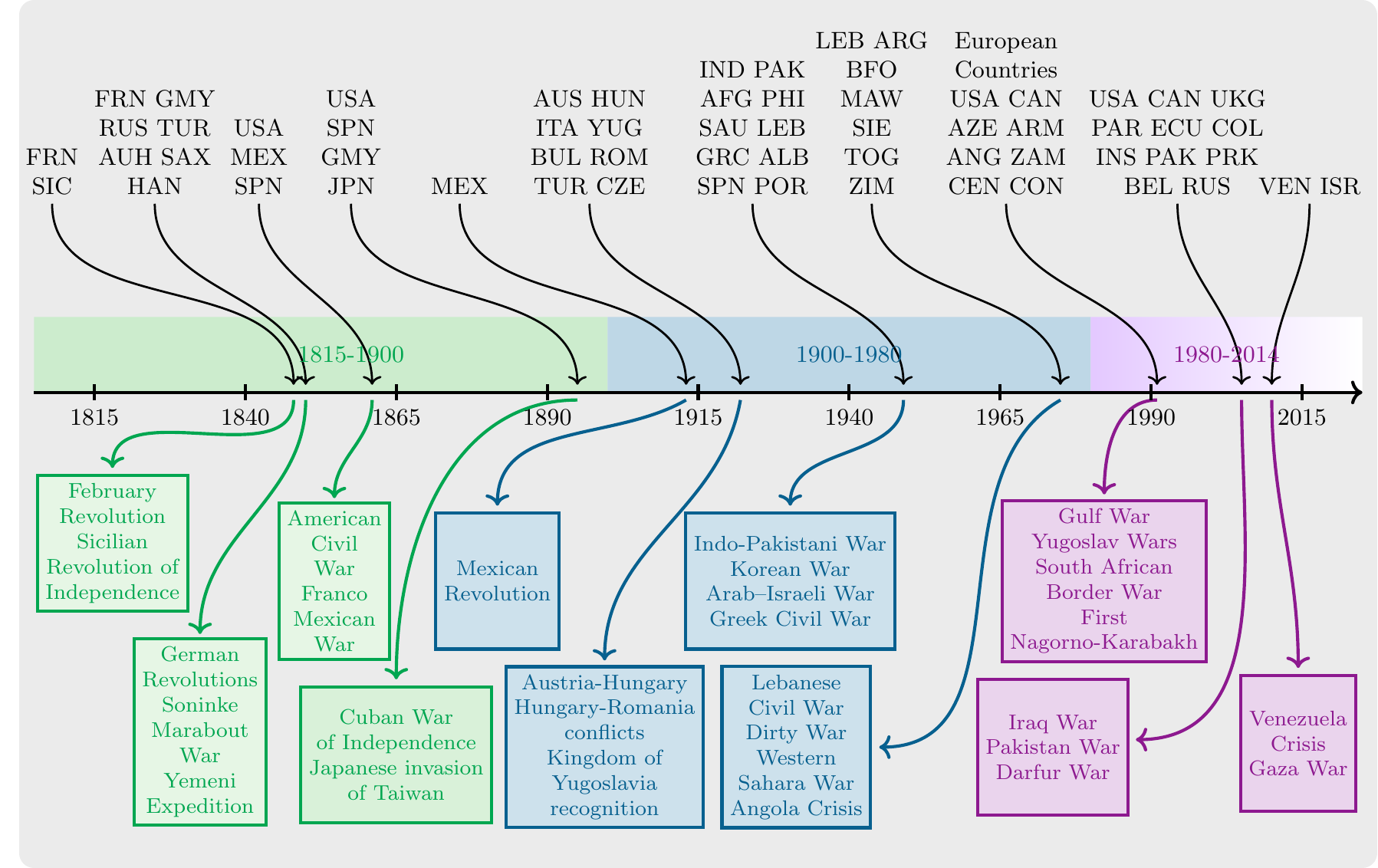} 
		\par\end{centering}
	\caption{Timeline of the major events corresponding to valleys detected in
		the XIX, XX and XXI centuries. A table relating each country to its corresponding code is provided in the SM.}
	\label{fig8} 
\end{figure}

\begin{figure}[h]
	\begin{centering}
		\includegraphics[width=0.85\textwidth]{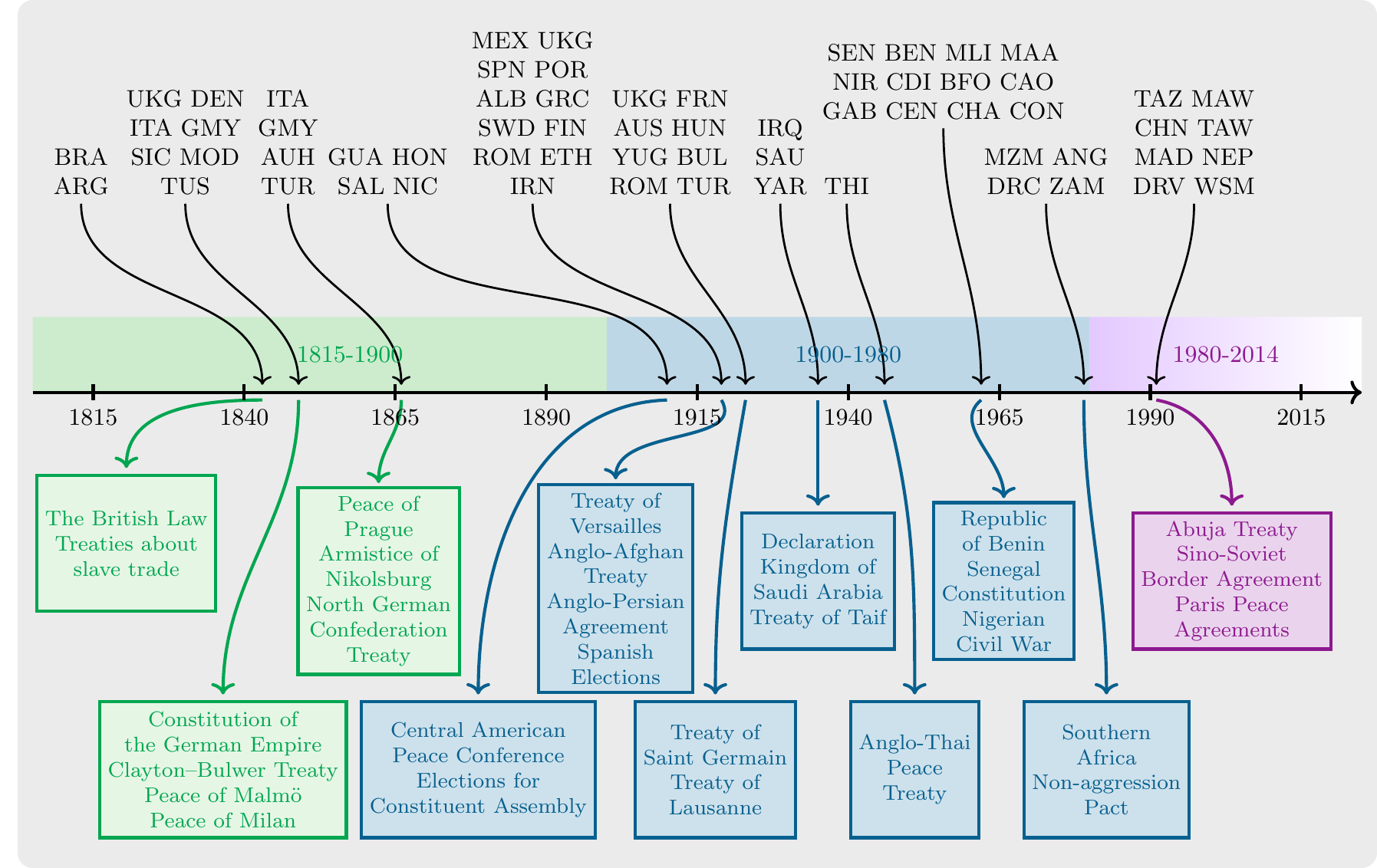} 
		\par\end{centering}
	\caption{Timeline of the major events corresponding to peaks detected in the XIX-XXI centuries. A table relating each country to its corresponding
		code is provided in the SM.}
	\label{increment} 
\end{figure}

Now that the statistical significance of the current method is confirmed,
we can perform an \textit{a posteriori} analysis of the matching between peaks
and valleys in terms of the nature of the historic events. In Table \ref{table_valleys} in the SM, we provide 12 examples of sharp drops in local balance,
which include for instance a drop of 0.138 in 1848 for France due
to the European revolutions taking place this year or a big drop of
0.422 in 1916 for Mexico due to the Mexican revolution. It is clear
that the drop in balance happens exactly at the time when a conflicting
historical event is taking place in the analyzed country. Similarly,
in Table \ref{table_peaks} in the SM, we give 12 instances of local balance increases,
which include for instance a big peak of 0.417 for Nigeria in 1963
due to the creation of the first Nigerian republic or a peak of 0.407
for Austria-Hungary in 1866 due to the Armistice of Nikolsburg and
the Peace of Prague reached this year. Again, we observe a match
between the sharp increase in balance and the involvement of the country
in treaties, protocols or independence events. In Fig. \ref{fig8} we illustrate a timeline containing the main events corresponding to valleys detected by the current methodology in the XIX, XX and XXI century, while in Fig. \ref{increment} we gather the main events that correspond to peaks. In the SM, we include more detailed figures illustrating timelines for each of the centuries analyzed here (figures \ref{timeline1}, \ref{timeline2} and \ref{timeline3}).

However, we still need to connect the quantitative analysis with a
qualitative narrative of the historic events happening in those countries.
To do so, we use here a ``mixed method'' based on the principles
described in \cite{mixed-1,mixed_methods}. Specifically, we retrospectively
analyze the historical situation in which the individual countries
were involved that year to create a narrative. This analysis reveals
some well-known major historic events and their main players, but
also some more subtle historic situations.

\subsubsection{Historic events associated with valleys}

As a first example that allows to bridge these quantitative findings with
a historic narrative, we will focus on the period 1848-1850, which
witnessed the largest and most widespread revolutionary wave in the
history of Europe. Historians agree in attributing the February 1848
Revolution in France, which led to the collapse of Monarchy, as the
spark that triggered the fire across Europe \cite{Evans2000}. The
node of France in the IR network is characterized by low values of
$\kappa_{v}$ in its whole recent history, but in the years preceding
the revolution of 1848, the value of $\kappa_{v}$ was around $0.40$.
In 1848 it drops to $0.25$ and in 1850 to $0.13$. In the meanwhile,
the uprisings spread throughout Europe, in particular in the Italian
and German states. In the early months of 1848, the Sicilian Revolution
took place and the value of $\kappa_{v}$ associated with the no longer
existing Kingdom of the Two Sicilies plunges from a value of $0.96$
in 1847 to a value of $0.22$ in 1848. 

In 1849 there is a general increase in the $\kappa_{v}$ values for
all states then in existence, but again in 1850 we see a new crash.
In fact, the German revolutions of 1848-1849 attempted to transform
the Confederation into a unified German Federal State; but in 1850
the Diet was re-established after the revolution was crushed by Austria
and Prussia \cite{Hahn2001}. This is observed by the sharp drops
in the values $\kappa_{v}$ of many German States: the Kingdoms of
Saxony and of Hanover both drop from $0.97$ in 1849 to $0.41$ in
1850, the Kingdoms of Bavaria and of Wurttenberg from $0.70$ to $0.38$
and the Grand Duchy of Baden from $0.70$ to $0.38$. Surprisingly,
the links of Hanover, Saxony and the Grand Duchy of Baden in 1850,
are all positive! And other German countries involved in such revolutions
have at most one negative edge only. Therefore, the local balance
index manages to capture the situation of local instability much better
than other local network indicators, such as the number of negative
edges incident to every country.

\begin{figure}[h]
	\begin{centering}
		\subfloat[]{\includegraphics[width=0.45\textwidth]{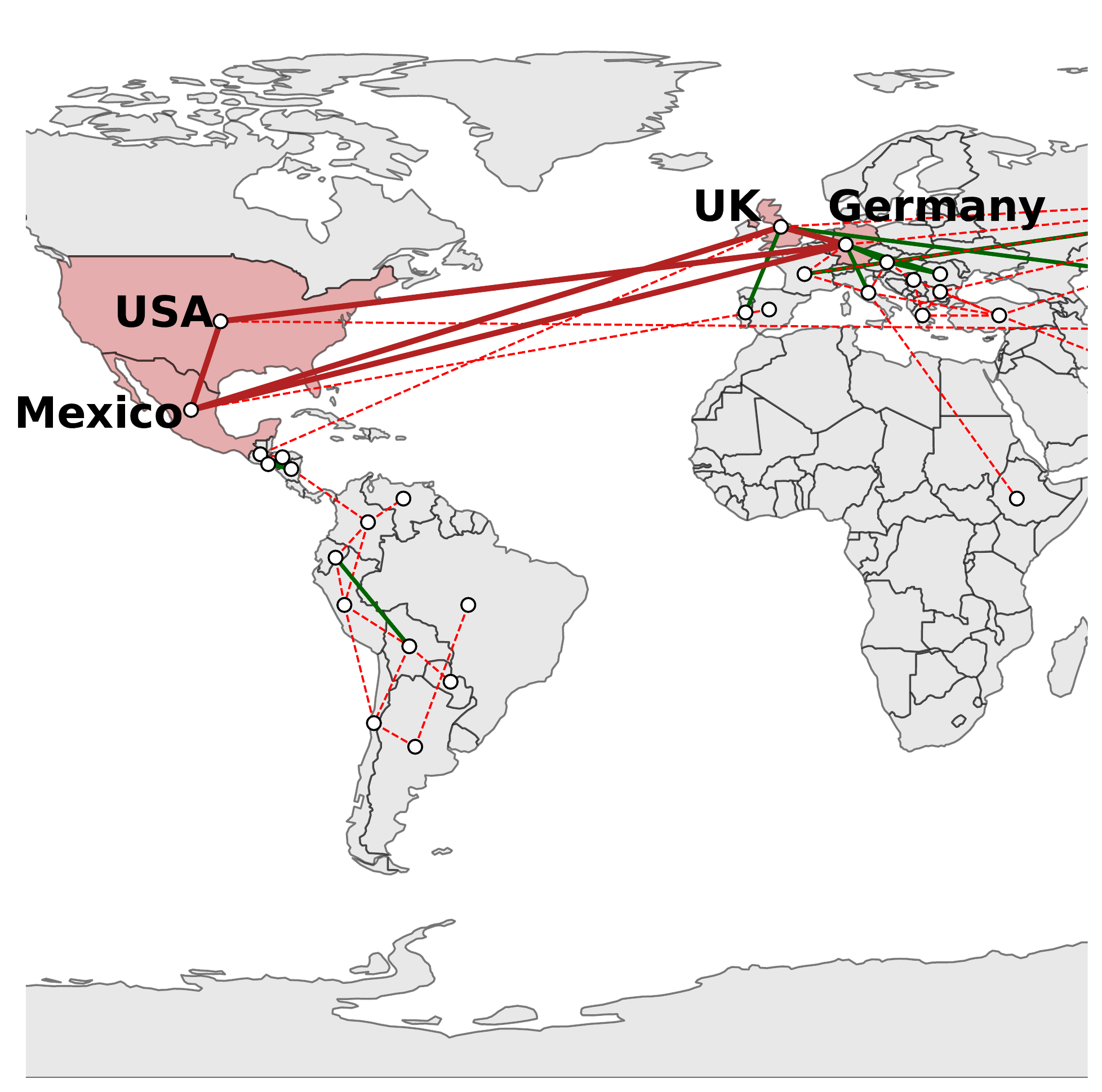}
			
		}\subfloat[]{\includegraphics[width=0.45\textwidth]{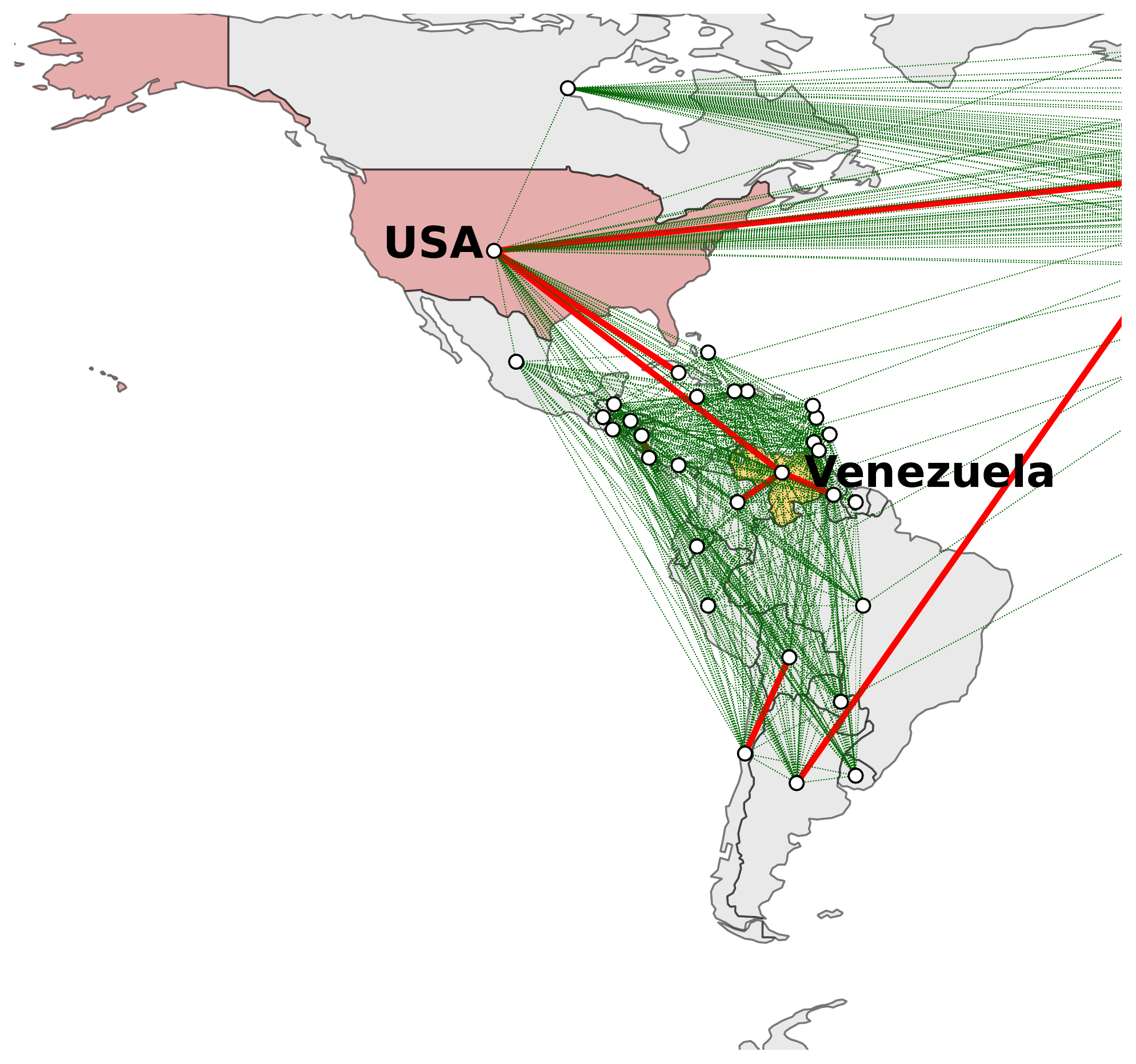}
			
		}
		\par\end{centering}
	\caption{Illustration of the IR involving (a) Mexico in 1913 and (b) Venezuela
		in 2010. In this map and all the following ones, green edges represent
		alliances and red edges depict conflicts or tensions.}
	\label{Mexico_Venezuela} 
\end{figure}

We now pick an example connecting our approach with historic narrative from the XX century.
A dramatic drop in the value of the local balance index is witnessed
by Mexico in 1913 and 1917. The $\kappa_{v}$ values for Mexico in
the early years of the XX century averaged around $0.96$. However,
in 1913 this value suddenly collapsed to $0.44$ and to $0.37$ in
1917. In 1912 Mexico's only link to the global network is through
the U.S., which is a negative edge. In 1913 the IR of Mexico are described
by the subgraph illustrated in Fig. \ref{Mexico_Venezuela}(a). The
majority of the edges in Fig. \ref{Mexico_Venezuela}(a) are negative.
The imbalance observed here is predominantly produced by the two negative
triangles Mexico-USA-Germany and Mexico-UK-Germany. Therefore, there
is a key role played by Germany in unbalancing the IR concerning Mexico
in 1913. If the Germany-Mexico link were positive, the balance of
this subgraph would significantly increase. The connection between
Germany and Mexico during this period has been ``relatively unnoticed
by historians until recently'' \cite{Leffler}. The narrative of
events is detailed in Sec. 3 of the SM and mainly refers to the efforts of Germany to draw Mexico into war with the United States with main goal of distract the last from an eventual European war.

Finally, we consider a historic scenario taking place in the XXI century. It refers to the Venezuelan economic crisis and how this destabilized
its IR by dropping dramatically its local balance. Venezuela exhibits
a very low local balance for the entire period, from the end of World
War II to the present, with an average of $0.25$. Nevertheless, in
2009 its value was $0.40$, followed by a dramatic drop the year after
to a value of $0.008$, one of the lowest values ever recorded for
all countries and for all years analyzed. In 2009, Venezuela was well
established in the South American cluster with only two negative edges,
one with Colombia and another with Guyana, and, very importantly,
with a positive edge with the U.S. In 2009 the link with the U.S.
represented for Venezuela a bottleneck through which it connected
to the rest of the world, outside South America. Now, in 2010 it happens
that this link suddenly becomes negative as can be seen in Fig. \ref{Mexico_Venezuela}(b),
and this is the only difference with the previous years. All the edges
within the South American cluster preserve their sign from 2009 to 2010. The events that sparked this situation are described in Sec. 3 of the SM and basically account for the economic crisis occurring in Venezuela in 2009-2010 and the subsequent deterioration of internal social situation which triggered sanctions from the European Union and United States.

\subsubsection{Historic events associated with peaks}

Let us now perform an analysis for peaks similar to the one done before for valleys.
We observe that events corresponding to peaks are more frequent in
the XIX and XX centuries, with practically no one in the XXI century.
Some of the detected events corresponding to peaks are illustrated
in Fig. \ref{increment}.


One of the historic periods showing significant incremental peaks in the local balance is the XIX century in South America. This was a very convulse period, which started with the independence of several of these regions from the Spanish and Portuguese empires, forming new states. It continues with the fights to balance power in the region, which also involves some global powers like Great Britain, France and Portugal. The local balance of Brazil increased from 0.58 in the previous year to 0.96, due to the removal of the negative edge of Brazil with Great Britain (maps in Fig. \ref{fig:Brazil} in the SM. Prior to 1843, Brazil had a clear confrontation towards Great Britain due to the British efforts to suppress the slave traffic \cite{Adams, Campbell}. In 1843 Argentina and Chile enter in conflict over Patagonia and the Strait of Magellan \cite{Burr} adding volatility to the region. Argentina drops again its local balance in 1845-46, possibly due to its conflict with France and Great Britain. Only after 1847 the local balance of Argentina is recovered to $\kappa_{i}\approx1$ due to negotiations to end the hostilities. Both Argentina and Brazil had territorial ambitions over Paraguay and Uruguay. Thus, a strategic alliance of Paraguay with Brazil against Argentina helped to increase the former's balance index in the South American network in the transition from 1850 to 1851. A similar strategic alliance with Peru allowed Venezuela to increase its balance from 0.42 in 1858 to 0.88 a year later \cite{Kroupa}.

Another dramatic increase of $\kappa_{v}$ observed in this work
occurred for Mozambique in 1977 and Angola in 1979. Mozambique had
$\kappa_{v}\approx0.008$ in 1976 and Angola had $\kappa_{v}\approx0.005$
in 1978. They become $0.602$ and $0.616$, respectively, in 1977
and 1979, respectively. The events most likely associated with this
increase in the local balance are the Angolan and the Mozambican civil
wars. In Sec. 3 of the SM, we describe the narrative of these wars which were developed in a Cold War scenario in which a perfectly balanced situation of two-blocs occurs in southern Africa with the participation of many countries around the world.

\subsection{Analysis of local balance time series}

\begin{figure}
	\includegraphics[width=\linewidth]{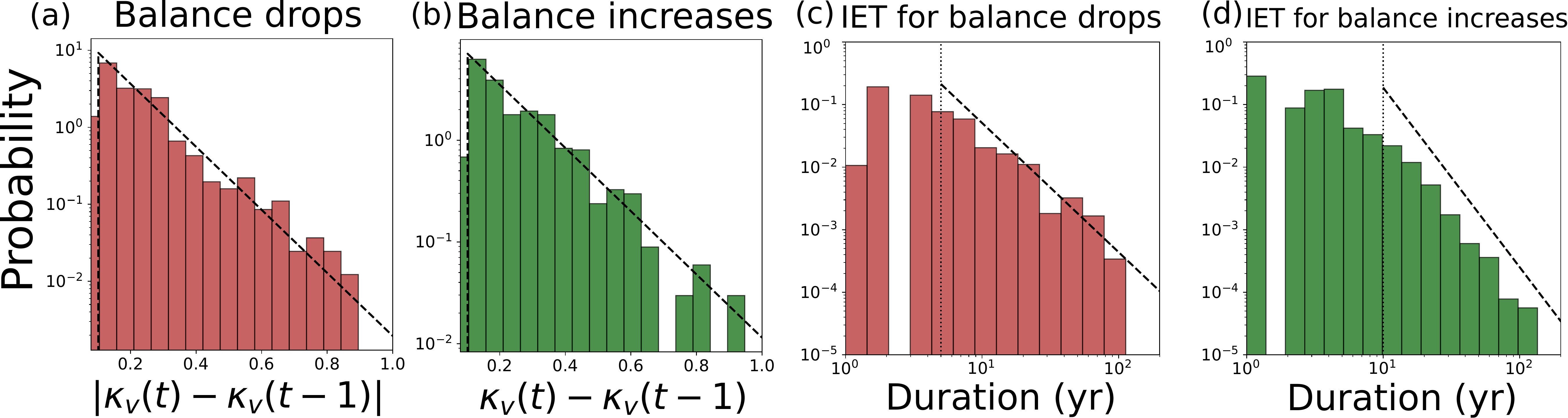}
	\caption{Probability distribution of the size of the local balance peaks
		(panel a) and valleys (panel b) and of the inter-event times (IET) between two
		consecutive peaks (panel c) and between two consecutive
		valleys (panel d). We also plot the best fit
		to each empirical distribution (dashed lines).The dotted vertical line indicates the $x_{min}$ after which we fit to a power-law distribution.}
	\label{fig:hist}
\end{figure}

\begin{figure}[H]
	\begin{centering}
		\centering \includegraphics[width=1\linewidth]{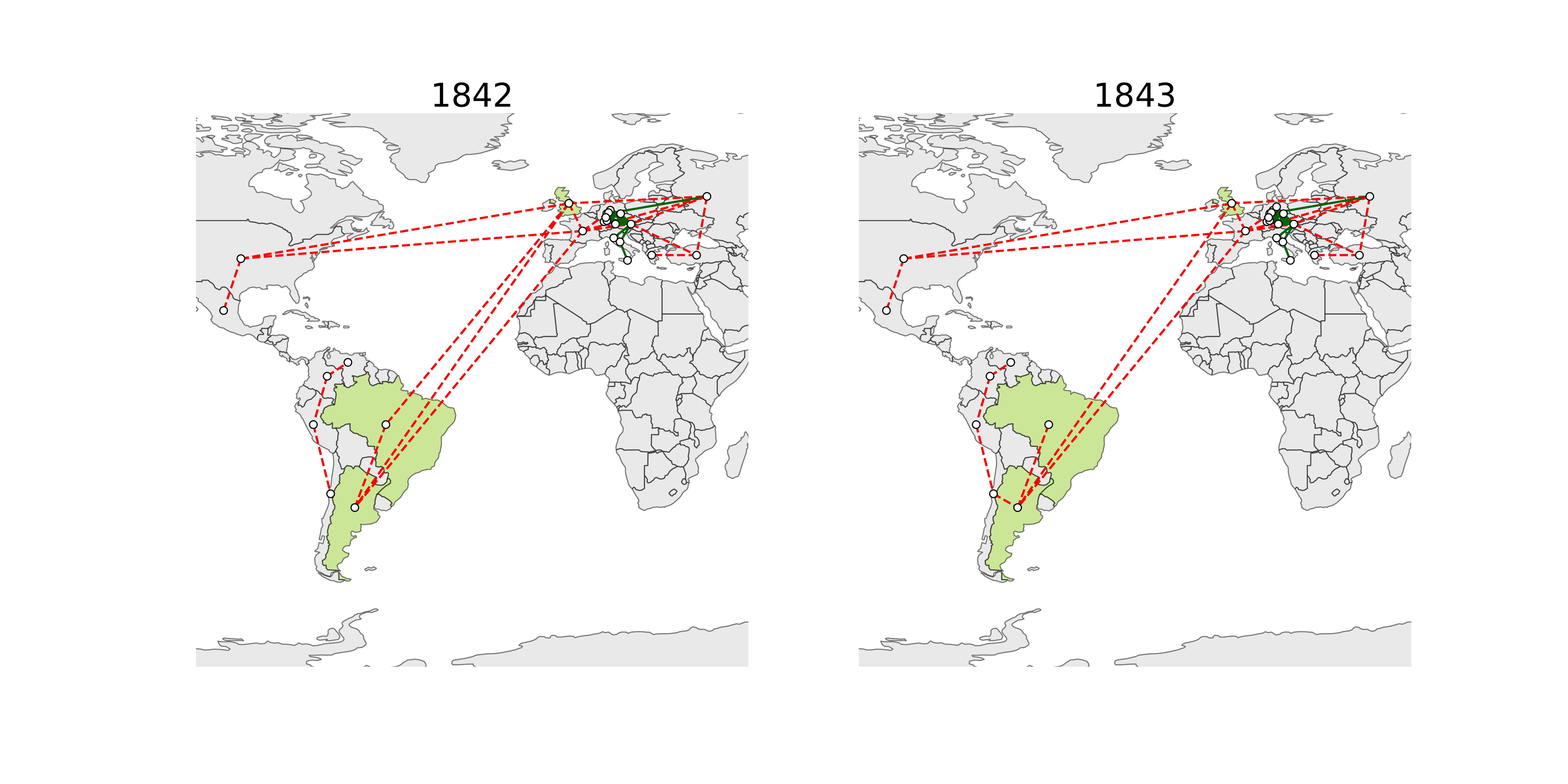}
		\par\end{centering}
	\caption{International relations network during the years 1842 and 1843. The
		main countries responsible for Brazil's increase in balance (Brazil,
		Argentina and the United Kingdom) are highlighted in green.}
	\label{fig:Brazil}
\end{figure}

This section is dedicated to extracting empirical probability distributions
from the time series of local balances for all countries. Specifically,
we analyze two magnitudes that encapsulate key properties of the balance
time series: the size of the balance drops and increases (i.e., the
probability that the local balance experiences a drop or increase
of size $x$, represented as $P(x):=P[\kappa_{v}(t)-\kappa_{v}(t-1)=x]$);
and the inter-event time (IET) distribution for drops and increases
(i.e., the probability that two consecutive drops/increases are separated
by a time interval $t$). The analysis is based on the extraction
of histograms that describe each distribution, followed by the identification
of the most plausible probability distribution and an estimation of
its parameters. To estimate the exponent of power-law distributions,
we have used the maximum likelihood estimation algorithm implemented
in the powerlaw python package \cite{powerlaw}. We have limited the
range of the power law distribution to the heavy-tail region; in other
words, we have imposed that $x>x_{min}$, where $x_{min}$ is another
estimated parameter. To compare the goodness of fit of the exponential
and power-law distributions, we have employed the Kolmogorov-Smirnov
(KS) statistical test. Since we are interested in capturing the statistics
of historically relevant events, we have filtered out of the size
distribution any event for which $|\kappa_{v}(t)-\kappa_{v}(t-1)|<0.1$.
The results of this analysis are presented in Fig. SM7 and \ref{fig:hist}.

In Fig. \ref{fig:hist}(a)-(b), we illustrate the frequency of balance drops
and increases as a function of their size. Both histograms represent
an exponential distribution: the distribution of balance increases
follows an exponential decay of the form $P(x)=\exp[-7.16(x-0.1)]$
(p-value: $p<10^{-6}$); while the distribution of balance drops decays
as $P(x)=\exp[-9.43(x-0.1)]$ ($p<10^{-4}$). One possible mechanism
that could generate such distributions is a scaled Poisson random process
in which random events happen at a constant rate $\alpha$. Indeed,
it is well-known that the inter-event time distribution of any scaled Poisson
process is given by $P(x)=\alpha e^{-\alpha x}$. This suggests that
the mechanisms that determine the size of a given historical drop
or increase event are not very different from a scaled Poisson random process.

In contrast, the distribution of inter-event times, shown in \ref{fig:hist}(c)-(d), does not follow an exponential distribution. Instead,
we find that power-law distributions fit the data more precisely:
for the IET of balance increases, the data is best fit to $P(t)\sim t^{-2.9}$
(p-value: $p=0.004$, $x_{min}=10$); while for the balance drops
IET we fit the histogram to $P(t)\sim t^{-2.1}$ ($p<10^{-4}$, $x_{min}=5$),
where $t$ is expressed in years. The estimated exponents for both
fits are smaller than three, meaning that the estimated probability
distributions have infinite variance. This suggests the existence
of a "scale-free" IET distribution, similar to those found in several other socio-technical systems \cite{Barabasi2005,Karsai2011}.
Significant events are not observable in the
XXI century, possibly because the ``noise'' level of the local balance
(continuous increments and drops) makes them statistically not relevant.

\subsubsection{Correlations in local balance time series}

We now examine the correlations between the behavior of different
countries in terms of local balance over time. The results are graphically
illustrated in Fig. \ref{Correlograms} in Sec. 3 of the SM. In panel (a), we show the Pearson correlation plot for the 'spatial distribution' of the $\kappa_{v}(t)$,
i.e. for pairs of different years, computed on the subset of countries
actually existing in both the years examined. In panel (b) we show
the Pearson correlation plot for the time series $\kappa_{v}(t)$
for pairs of countries, computed on the intersection of their corresponding
existence time intervals.

The correlogram in panel (a) reveals a clear historical caesura corresponding
to the year 1949, after the Second World War: the years before 1949
are predominantly positively correlated with each other, as are the
years after 1949; but the years in the two blocs are typically negatively
correlated. This implies a noticeably different distribution of the
local balance among countries before and after that year.

On the other hand, the correlogram in panel (b) shows a positive correlation
in the time evolution of the individual balance of most countries.
In fact, throughout the whole set of nations only $14.38\%$ of the
pairs of nations are negative correlated. Among the nations with the
highest number of negative correlations we find many African countries.
For example, Kenya is anticorrelated with 83 countries, Mauritania
with $82$, and Mali and Nigeria with $76$. On the other hand, among
the nations with the highest number of positive correlations we find
Ukraine, Belarus, Armenia, Azerbaijan and Kazakhstan with only one
anticorrelated country.

If we instead ask ourselves which pair of countries shows the largest
correlation throughout their time series, we find that this pair is
comprised of Spain and Portugal, with a value of the correlation coefficient $\rho=0.899$. Another
highly correlated pair of countries is France and the United Kingdom
($\rho=0.765$). It is worth noting that the last major conflict between
United Kingdom and France was that of the Napoleonic Wars (1793-1815),
which finished one year before the starting year of the data set.
Other pairs of maximally correlated countries are Russia with France
($\rho=0.746$), Germany with Italy ($\rho=0.759$), Italy with Turkey
($\rho=0.799$), and Yugoslavia with Romania ($\rho=0.806$).

One important global trend that can be acknowledged is the growth
of correlation within specific clusters of nations in the second half
of the XX century, after World War II. For instance, let us consider
the founding member countries of NATO, on the one hand, and the Warsaw
Pact, on the other. The mean internal correlation between the values
of the local $\kappa_{v}(t)$ in the period 1945-2014 is $\rho=0.890$
for the NATO cluster and $\rho=0.899$ for the Warsaw Pact cluster.
If we consider the whole interval 1816-2014, the mean correlations
within the two clusters are $\rho=0.635$ and $\rho=0.835$, respectively.
Thus, we conclude that NATO has contributed much more to changing
the \textit{status quo} than the Warsaw Pact, which just confirmed
a pre-existing system of consistent behaviors. Let us remark that the role of supranational coalitions, such as NATO and the former Warsaw Pact, in stabilizing cooperation, reducing instability, and creating homogeneous behavior among coalition members has recently been highlighted, for example, in \cite{Galam2023}.

\section{Conclusion}\label{Conclusion}
With this work we have been aimed at demonstrating how mathematics can
help to reveal that ``not everything in history is contingent and
particular'' \cite{cliodynamics_1}. We do so by using algebraic
graph theory techniques, which combined with appropriate data analysis
tools reveal patterns in history from the individual participation
of countries in the global web of IR. For instance, the current work
informs that the probability that two events are separated by a time
$t$ is best fit to the power law $p\left(t\right)\sim t^{-2.9}$.
This means that while the probability that such events are separated
by only 1-2 years is about 70\%, it drops to nearly 10\% for intervals
of 2-3 years and it is almost negligible for separations longer than
5 years. We have found that the main reason why such events occur
so frequently is because they are populated by different countries.
Indeed, if $p\left(N\right)$ is the probability that a given country
is involved in $N$ major events dropping the balance in the whole
period of 199 years, then $p\left(N\right)\sim N^{-1.405}$. This
means that the probability that a given country is involved in only
one of such events is about 42\%, but it drops to 16\% for being involved
in two events in the whole period. The probability that one single
country is involved in 10 events in 199 years is only 2\%. But history
makes exceptions! Chile and Argentina were involved in 15 events,
Peru in 16, Brazil in 19 and Spain in 20, mainly due to the convulsive
situation existing in South America in the XIX century analyzed before.
To sum up, the local balance index captures information about how
an individual country reacts to the change of balance of the rest
of states in the world at a given time. Therefore, the local balance
index can be thought of as a quantitative candidate for translating
mathematically what historians and political scientists call the ``balance
of power''.

\backmatter


\bmhead{Acknowledgements}

EE thanks Zeev Maoz for sharing data and for discussions at an initial stage of this work.\\
FDD and EE acknowledge funding from the Spanish Ministerio de Ciencia e Innovación, Agencia Estatal de Investigación Program for Units of Excellences María de Maeztu (CEX2021-001164-M /10.13039/501100011033).  FDD thanks financial support MDM-2017-0711-20-2 funded by MCIN/AEI/10.13039/501100011033 and by FSE invierte en tu futuro, as well as project APASOS (No. PID2021-122256NB-C22). EE also acknowledges funding from project OLGRA (PID2019-107603GB-I00) funded by Spanish Ministry of Science and Innovation.

%
%


%
%
%
%

\newpage
\setcounter{section}{0}

\section*{Supplementary Material of the paper Mathematical Modeling of Local Balance in Signed Networks and Its Applications to Global International Analysis}
\section{Proof of Propositions \ref{signedcycles} and \ref{negativeclique}}
\label{appendixA}
\setcounter{theorem}{4}
\begin{proposition}
	Let $C_{n}^{k-}$ be a signed cycle of length $n$ and $k$ negative
	edges. Then, the local balance index of any node in $C_{n}^{k-}$
	is
	
	\begin{equation}
		\kappa_v\left(C_{2l}^{k-}\right)= \begin{cases}
			\dfrac{\sum_{j=0}^{2l-1}\exp\left({2\cos\left(\dfrac{\left(2j+1\right)\pi}{2l}\right)}\right)}{\sum_{j=0}^{2l-1}\exp\left({2\cos\left(\dfrac{j\pi}{l}\right)}\right)} & k\textnormal{ odd},\\
			1 & k\textnormal{ even},
		\end{cases}
	\end{equation}
	
	for even cycles and
	
	\begin{equation}
		\kappa_v\left(C_{2l+1}^{k-}\right)= \begin{cases}
			\dfrac{\sum_{j=0}^{2l}\exp\left({2\cos\left(\dfrac{2j\pi}{2l+1}\right)}\right)}{\sum_{j=0}^{2l}\exp\left({2\cos\left(\dfrac{\left(2j+1\right)\pi}{2l+1}\right)}\right)} & k\textnormal{ odd},\\
			1 & k\textnormal{ even},
		\end{cases}
	\end{equation}
	
	for odd ones.
\end{proposition}
\begin{proof}
	In a signed cycle graph the number of closed walks of a given length
	are the same for every node in the cycle. Then, the local balance
	is
	\begin{align*}
		\kappa_v\left(C_{n}^{k-}\right)=\frac{(e^A)_{vv}}{(e^{|A|})_{vv}} = \frac{\dfrac{1}{n}\sum_{j=1}^{n}e^{\alpha_{j}\left(C_{n}^{k-}\right)}}{\dfrac{1}{n}\sum_{j=1}^{n}e^{\beta_{j}\left(C_{n}\right)}},
	\end{align*}
	where $\{\alpha_j\}$ are the eigenvalues of $C_n^{k-}$ and $\{\beta_j\}$ are those of $C_n$.
	
	It is known that any closed walk in a cycle with an even number of negative
	edges will always be positive; therefore,
	the local balance index is unity for any node. That is, $\kappa_v\left(C_{n}^{k-}\right)=1$
	if $k$ is even. Additionally, all signed cycles with equal number of
	nodes and the same parity in the number of negative edges share the same spectrum. In particular, as proved before \cite{akbari2018spectral}, the spectrum of signed cycles with odd length and any odd number of negative edges is:
	
	\begin{equation*}
		Sp\left(C_{2l}^{k-}\right)=\left\{ 2\cos\left(\dfrac{\left(2j+1\right)\pi}{2l}\right),\:j=0,1,\ldots,2m-1\right\} ,
	\end{equation*}
	
	\begin{equation*}
		Sp\left(C_{2l+1}^{k-}\right)=\left\{ 2\cos\left(\dfrac{2j\pi}{2l+1}\right),\:j=0,1,\ldots,2m\right\} .
	\end{equation*}
	
	These expressions combined with those for the spectra of the unsigned
	cycles:
	
	\begin{equation*}
		Sp\left(C_{2l}\right)=\left\{ 2\cos\left(\dfrac{2j\pi}{2l}\right),\:j=0,1,\ldots,2m-1\right\} ,
	\end{equation*}
	
	\begin{equation*}
		Sp\left(C_{2l+1}\right)=\left\{ 2\cos\left(\dfrac{\left(2j+1\right)\pi}{2l+1}\right),\:j=0,1,\ldots,2m\right\} ,
	\end{equation*}
	
	give rise to the final proof.
\end{proof}

\begin{proposition}
	The local balance of a $K_{n}(K_{l}^{-})$ graph, $\kappa_{v}\left(K_{n}(K_{l}^{-})\right)$,
	tends to zero when $l\to\infty$ and $m:=n-l$ remains finite. 
\end{proposition}

\begin{proof}
	To find the global and local balances of a graph, the first step is
	to find the spectrum of its adjacency matrix. The adjacency matrix
	of a $K_{n}(K_{l}^{-})$ clique has the following form: 
	\begin{align*}
		A=\left(\begin{matrix}I_{ll}-E_{ll} & E_{lm}\\
			E_{ml} & E_{mm}-I_{mm}
		\end{matrix}\right),
	\end{align*}
	where $I$ is the identity, $E$ is a matrix of ones, and the subindices
	indicate the dimensions of each matrix. Immediately, we observe that
	the first $l$ rows of the matrix $A-I$ are equal. Therefore, $\alpha=1$
	is an eigenvalue with multiplicity $l-1$. In the same way, we find
	that $\alpha=-1$ is an eigenvalue with multiplicity $m-1$ because
	the $m$ last rows of $A+I$ are equal. \\
	To find the last two eigenvalues, we conjecture that the vector $\psi=(a\vec{1}_{l},\ b\vec{1}_{m})^{T}$
	is an eigenvector of $A$, and try to find the parameters $a$ and
	$b$. Under this Ansatz, the eigenvalue equation $A\psi=\alpha\psi$
	is fulfilled if and only if $mla-{(m-1-\alpha)(1-l-\alpha)}a=0$.
	Imposing $a\neq0$, we get the last two eigenvalues: 
	\begin{align*}
		\alpha_{\pm}=\frac{m-l\pm\sqrt{(m-l)^{2}+8ml-4l-4m+4}}{2}
	\end{align*}
	Notice that, since $2ml\geq m+l$ for $m,l\geq1$, the term inside
	the square root is always larger than $(m-l)$. Hence, $\alpha_{+}>2(m-l)$
	and $\alpha_{-}<0$. \\
	On the other hand, we also need the spectrum of $|A|$. Since $K_{n}(K_{l}^{-})$
	is a signed complete graph, the spectrum of its unsigned counterpart
	is $Sp(|A|)=\{n-1^{[1]},1^{[n-1]}\}$. The spectra of $A$ and $|A|$
	uniquely determine the global balance index: 
	\begin{align*}
		\kappa_{gl}(G)=\frac{\sum_{j}e^{\alpha_{j}}}{\sum_{j}e^{\beta_{j}}}=\frac{e^{\alpha_{+}}+e^{\alpha_{-}}+(m-1)e^{-1}+(l-1)e}{e^{n-1}+(n-1)e^{-1}}.
	\end{align*}
	Up to now, every calculation was valid for any $m,l$. Now, we make
	use of the limit $l\to\infty$. In this limit, we can Taylor expand
	the eigenvalues $\alpha_{\pm}$: 
	\begin{align*}
		\alpha_{\pm} & =\frac{m-l}{2}\left(1\mp\sqrt{1+\frac{8ml-4l-4m+4}{(m-l)^{2}}}\right) \nonumber \\
		& =\frac{m-l}{2}\left(1\mp\ \left(1+\frac{8ml-4l-4m+4}{2(m-l)^{2}}+O(l^{-3})\right)\right) \nonumber \\
		& =\frac{m-l}{2}\left(1\mp\left(1+\frac{4m-2}{l}+O(l^{-2})\right)\right);
	\end{align*}
	and specifically, 
	\begin{align*}
		\alpha_{+} & =(2m-1)+O(l^{-1}),\\
		\alpha_{-} & =(1-l-m)+O(l^{-1}).
	\end{align*}
	The key insight of this calculation is that none of the eigenvalues
	of $A$ \textit{increases} linearly with $l$. As a consequence, the global
	balance index decreases exponentially fast to zero as $l\to\infty$. \\
	To find the scaling of the local balance index, we first need the scaling
	of the subgraph centrality $SC_{v}:=(e^{|A|})_{vv}$. Since $|A|$
	is a complete graph, it is invariant under any permutation of the
	nodes, so $SC_{v}=\frac{tr(e^{|A|})}{n}=\frac{{e^{n-1}+(n-1)e^{-1}}}{n}$.
	Importantly, the subgraph centrality of every node diverges exponentially
	as $l$ (and thus $n$) tend to infinity. \\
	Now we are in a position to obtain the asymptotic behavior of the
	local balance index. To do so, we use the following identity: 
	\begin{align*}
		(e^{A})_{vv}=\sum_{v}\kappa_{v}(G)SC_{v}
	\end{align*}
	The left-hand side of this equation was already calculated:  $(e^{A})_{vv}=e^{\alpha_{+}}+e^{\alpha_{-}}+$ \\$ +(m-1)e^{-1}+(l-1)e$.
	As discussed before, none of the eigenvalues increase linearly with
	$l$, so $(e^{A})_{vv}$ scales as $O(l)$ as $l\to\infty$. Therefore,
	$\sum_{v}\kappa_{v}(G)SC_{v}$ must also scale as $O(l)$. But we
	previously calculated that the subgraph centrality scaled as $O(e^{n})$,
	or equivalently, since $m$ is finite, as $O(e^{l})$. The local balance
	term must compensate this exponential divergence by scaling as $O(e^{-l})$.
	But this means that, as $l\to\infty$, every local balance index $\kappa_{v}(G)$
	must tend towards zero; hence, the proposition is proved. 
\end{proof}

\section{Measures of matrix similarity}
\label{appendixB}
In Sec. 4.1.4, we used two measures to quantitatively measure the similarity between the ground truth matrix and the real and randomized local balance matrices. Here, we briefly describe these measures and some of their properties. \\
The fist similarity measure that we propose, called the Frobenius similarity, is defined by:
\begin{align}
	s_F(A,B)= \frac{tr(A^TB)}{\sqrt{tr(A^TA)tr(B^TB)}}   \label{eq:similarity2}
\end{align}
We recall that the operation $tr(A^TB)$ defines an inner product between two matrices $A$ and $B$, and therefore $\sqrt{tr(A^TA)}$ defines the so-called the Frobenius norm. Consequently, the quotient \eqref{eq:similarity2} equals one if and only if $A=B$. In fact, it is mathematically guaranteed that $s_F\in[-1,1]$, and $s_F = -1$ when $A=-B$. Another important property of the Frobenius similarity is that it is a particular instance of a cosine similarity measure, which are arguably one of the most popular families of similarity metrics. Finally, we note that our measure is applicable to non-square matrices, as long as $A$ and $B$ share the same dimensions.

The second similarity measure that we use is the Pearson correlation coefficient between the two (flattened) matrices. To compute it, we first convert the $N_c\times N_t$ matrices $A$ and $B$ into vectors $\tilde A, \tilde B$ of dimensions $N_c N_t$ obtained by concatenating the rows. Then, we apply the usual definition of the Pearson correlation to these vectors:
\begin{align}
	s_P(A,B) = \frac{Cov(\tilde A, \tilde B)}{\sqrt{Var(\tilde A)Var(\tilde B)}}
\end{align}
Since $s_P$ is a Pearson coefficient, it fulfills $s_P\in[-1,1]$, with $s_P=1$ when $A=B$ and $s_P=-1$ when $A=-B$.

In Table \ref{signed_correlations_table} we provide the measures of the two similarity coefficients between the ground truth matrix and the peaks and valleys matrix, as well as the similarity obtained for the randomized null model (see main text for details). 
\begin{table}[h]
	\begin{centering}
		\begin{tabular}{|c|c|c|}
			\hline 
			\multirow{1}{*}{} & real & rand\tabularnewline
			\hline 
			Pearson & 0.306  & 0.039 \tabularnewline
			\hline 
			Frobenius & 0.313  & 0.052 \tabularnewline
			\hline 
		\end{tabular}
		\par\end{centering}
	\caption{Similarity measures between the ground truth matrix of historical events and the matrix of peaks and valleys.  The first column indicates the similarity between the ground truth and the real matrix of peaks and valleys, $T$, while the second column provides the average similarity after 500 randomizations of the local balance time series, $T'$, as detailed in the main text.}
	\label{signed_correlations_table}
\end{table}

\section{Linking local balance with historical narrative}
\label{appendixC}

\subsection{The Mexican Revolution}
Let us discuss here the political events that happened in Mexico during
the period analyzed in the main text. In February 1913, during the "Ten Tragic Days",
a military coup ousted President Madero, with General Victoriano Huerta
taking his place as the new president \cite{Richmond}. Although this
revolutionary conflict was primarily a civil war, foreign powers and
private companies, having important economic and strategic interests
in Mexico, played a remarkable role \cite{Leffler}. On one side,
Germany was trying to deliver arms to the Huerta regime, to which
the U.S. was opposed and acted by the seizure and occupation of the
port of Veracruz to avoid the arming of Huerta's forces from the German
side \cite{Adolfo2006}. Since Huerta did not seize power until February
1913, Germany was during the beginning of that year an ally of a rebel
and therefore an enemy of the Mexican government. With the outbreak
of World War I in Europe, Germany attempted to draw Mexico into war
with the United States, which was itself neutral at the time. On the
other hand, Germany hoped to draw the U.S. troops from deployment
in Europe and, as a reward in the case of German victory, to return
to Mexico the territory that it lost in the Mexican-American War.
Therefore, the negative edge Germany-Mexico disappeared after 1913,
as Germany now conspired to inflame Mexico against the U.S. The leaking
of this information pushed the U.S. into war against Germany in 1917,
consequently dropping Mexico's balance index to $\kappa_{v}=0.37$.
The German activities in Mexico, although, were decisive for the events
surrounding the entry of the United States into the First World War.
These events show how, even without large-scale war interventions,
Germany played a key role in destabilizing Mexico and significantly
dropping its local balance indicator.

\subsection{Venezuela's economic crisis}
We will discuss here the historic events that led to the balanced shifts detected in Venezuela. The main trigger was a deep banking crisis in
the transition from 2009 to 2010 linked to episodes of corruption
in the government of Hugo Chavez. The second aspect is related to
an out-of-control rise in inflation and an unprecedented escalation
of starvation, disease, crime and high mortality rates, which have
led to mass emigration from the country and ongoing violations of
human rights. It is considered the most severe and dramatic crisis
ever experienced by a non-war-ridden country in recent times \cite{Bull}.
In response to this dramatic situation, the European Union, the United
States and other countries applied individual sanctions against government
officials and members of both the military and security forces \cite{Seelke}.
On 2 June 2010, Chavez declared the \textit{economic war}, due to
increasing shortages of food and basic necessities.
\begin{figure}[H]
	\centering \includegraphics[width=1\linewidth]{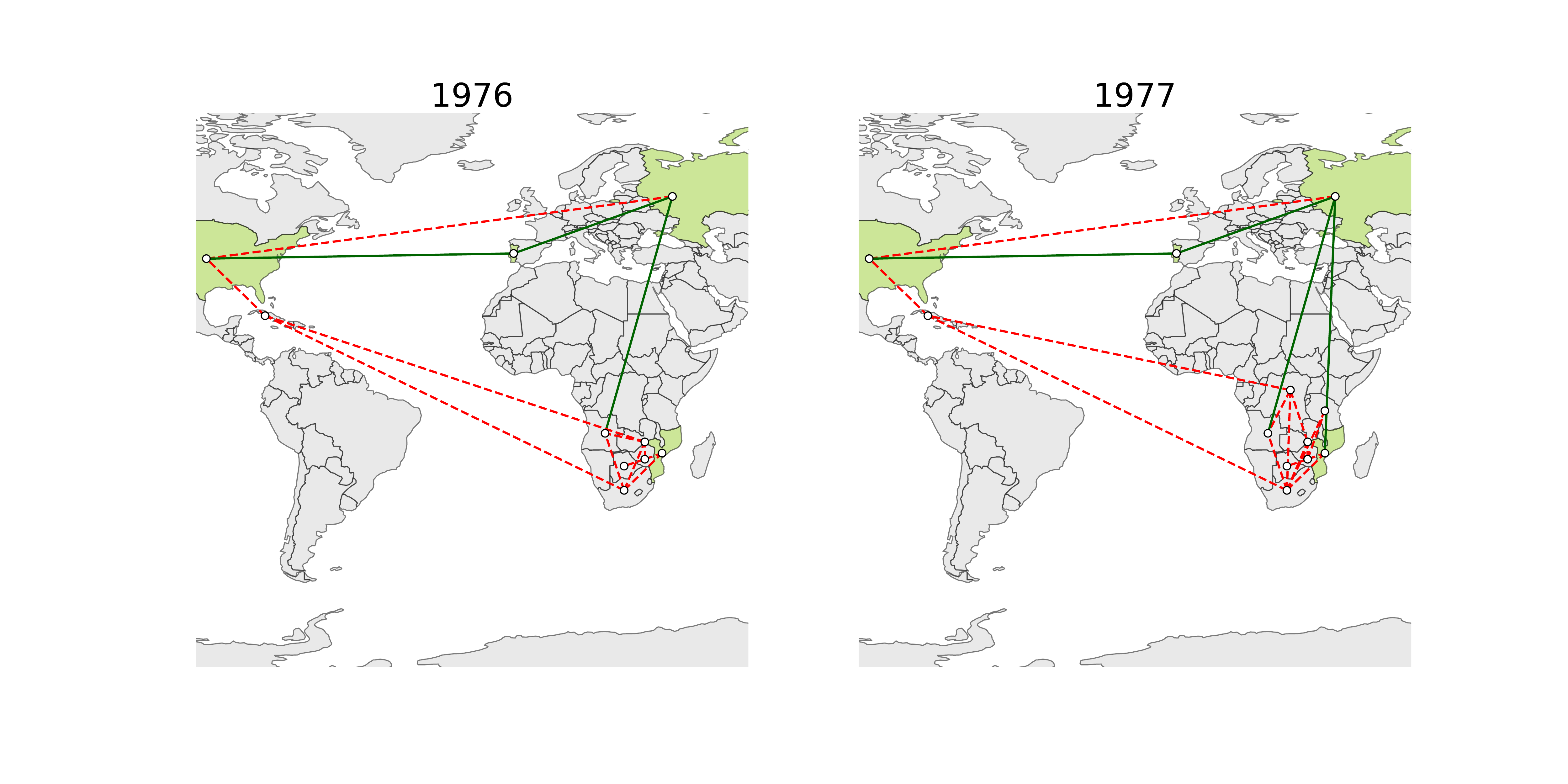}
	\caption{Network of international relations during Mozambique's independence
		(1976-1977). We highlight the countries playing a major role in Mozambique's
		independence: Mozambique, Portugal, the USA and the USSR. To improve
		clarity, we have only illustrated the subgraph induced by the mentioned
		four countries and Mozambique's first and second nearest neighbors.}
	\label{fig:Mozambique}
\end{figure}

\begin{figure}[H]
	\centering \includegraphics[width=1\linewidth]{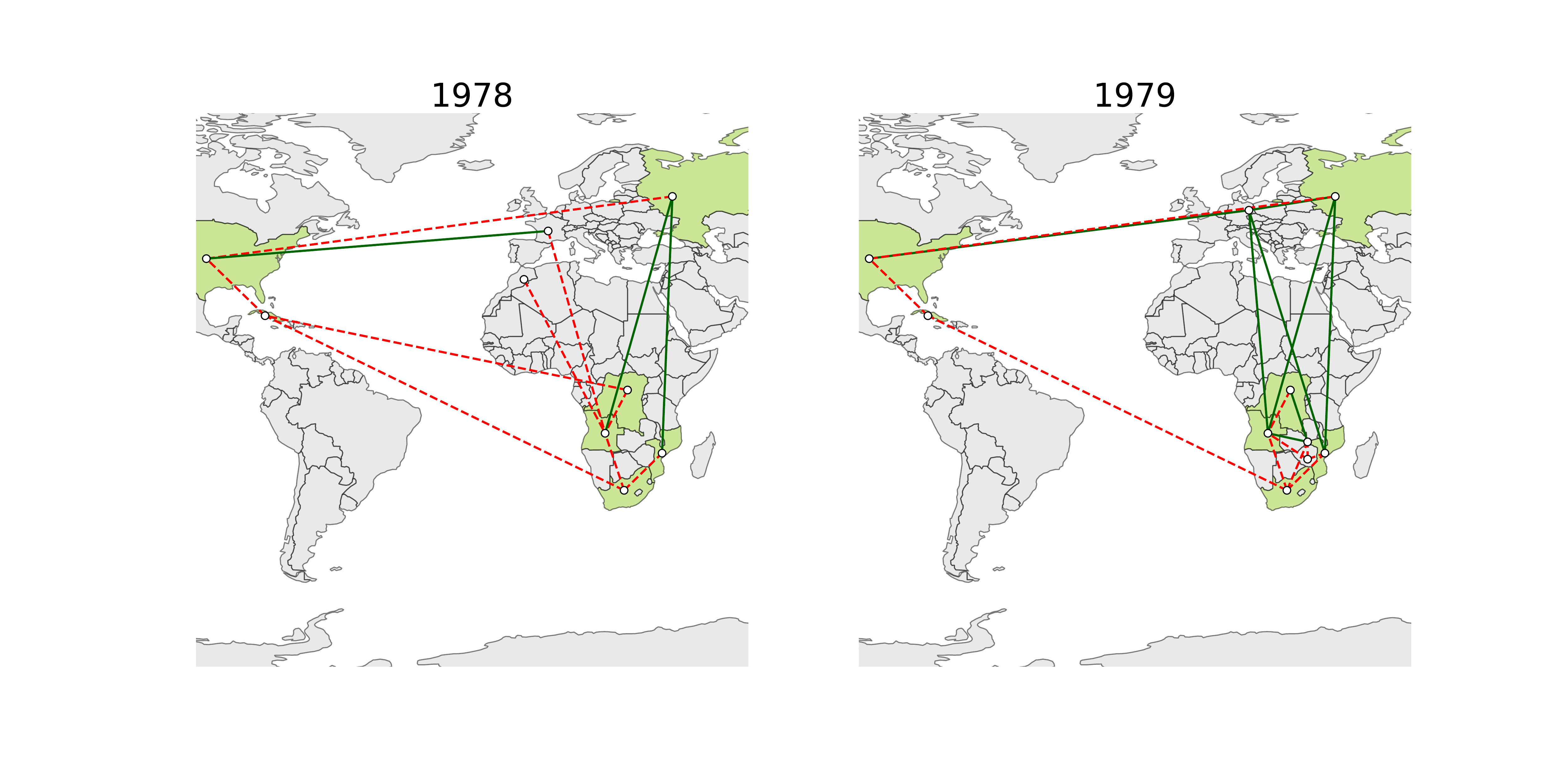} \caption{Network of international relations during Angola's independence (1978-1979).
		We highlight the countries involved in Angola's conflict: Angola,
		the United States of America, the USSR, Cuba, Mozambique, South Africa,
		and Zaire. To improve clarity, we have only illustrated the subgraph
		induced by these countries and Angola's first nearest neighbors.}
	\label{fig:Angola}
\end{figure}

\subsection{The Angolan and Mozambican civil wars}

In this subsection we will comment on the Angolan and Mozambican civil wars. Both wars can be seen as a consequence of the two-bloc structure
created during the Cold War (Figs. SM2 and SM3). For instance, in the
Mozambican conflict that erupted after the country gained independence
from Portugal, the Soviet Union aligned with the Mozambican government
in 1977, while the US supported the insurgents. More dramatic was
the situation after the independence of Angola from Portugal in 1975.
A war started between three main players in the country: the \textit{People's
	Movement for the Liberation of Angola} (MPLA), the \textit{National
	Union for the Total Independence of Angola} (UNITA) and the \textit{Front
	for the National Liberation of Angola} (FNLA). On the one side, MPLA
was supported by the Soviet Union, Cuba, and Mozambique, while on
the other, UNITA and FNLA were mainly supported by the U.S., Zaire
and South Africa. In 1978, Angola was a vertex of a balanced square
with two positive edges from Soviet Union to both Angola and Mozambique
and two negative ones to the same countries from South Africa. In
1979, the Non-Aggression Pact for Southern Africa was signed by the
presidents of Angola, Zaire and Zambia. Then, Angola formed a new
positive link with Zambia, which makes it take part in two balanced
triangles, each having one positive and two negative edges. Angola
also formed an unbalanced triangle with Zambia and Congo. This was
a long war that extended until 2002 with a large number of casualties,
wounded and refugees.

\begin{figure}
	\begin{centering}
		\includegraphics[width=0.9\textwidth]{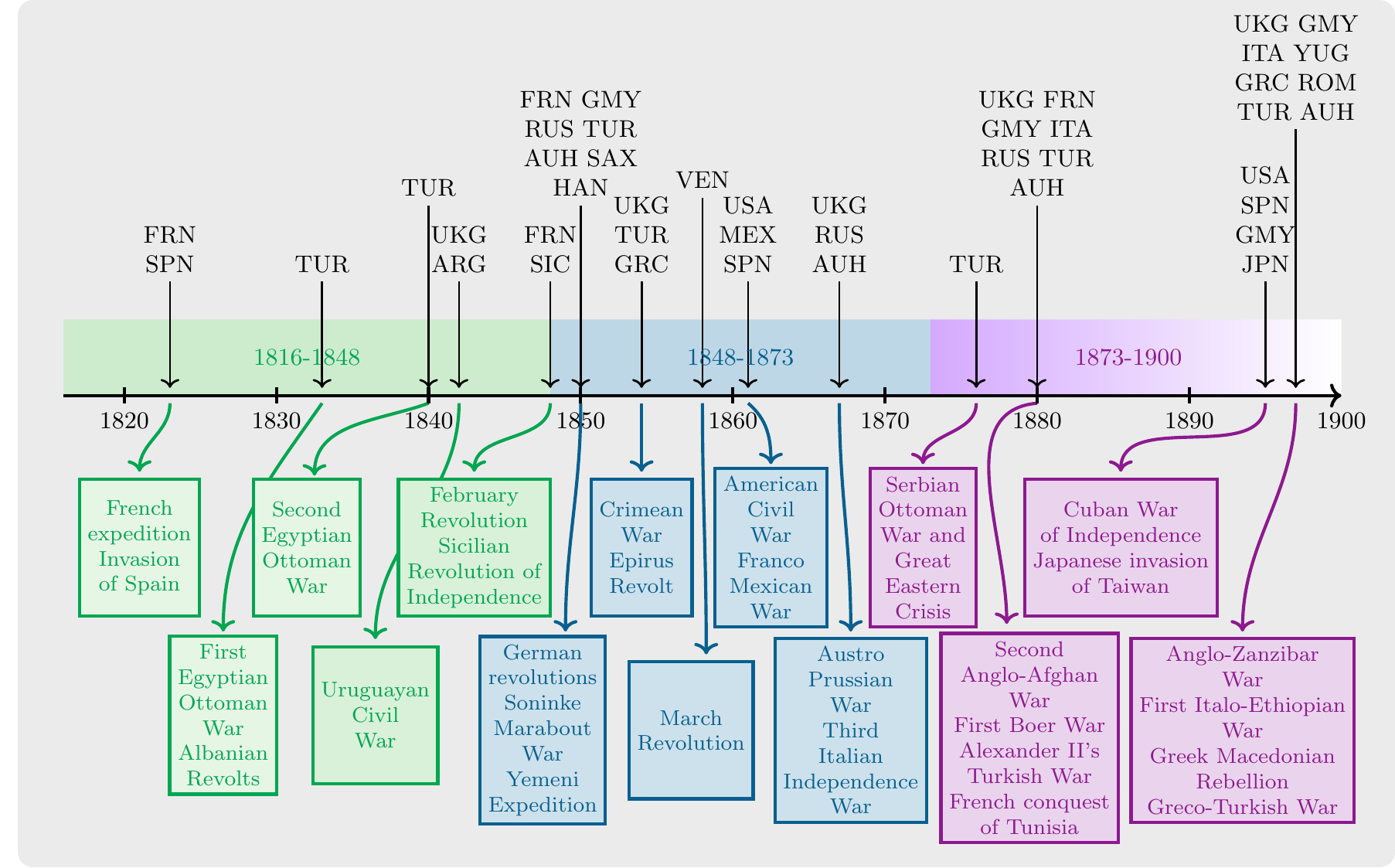}
		\par\end{centering}
	\caption{Timeline of the major events corresponding to valleys detected in
		the XIX century. A table relating each country to its corresponding
		code is provided in the SM \ref{appendixC}.}
	\label{timeline1}
\end{figure}

\begin{figure}
	\begin{centering}
		\includegraphics[width=1\textwidth]{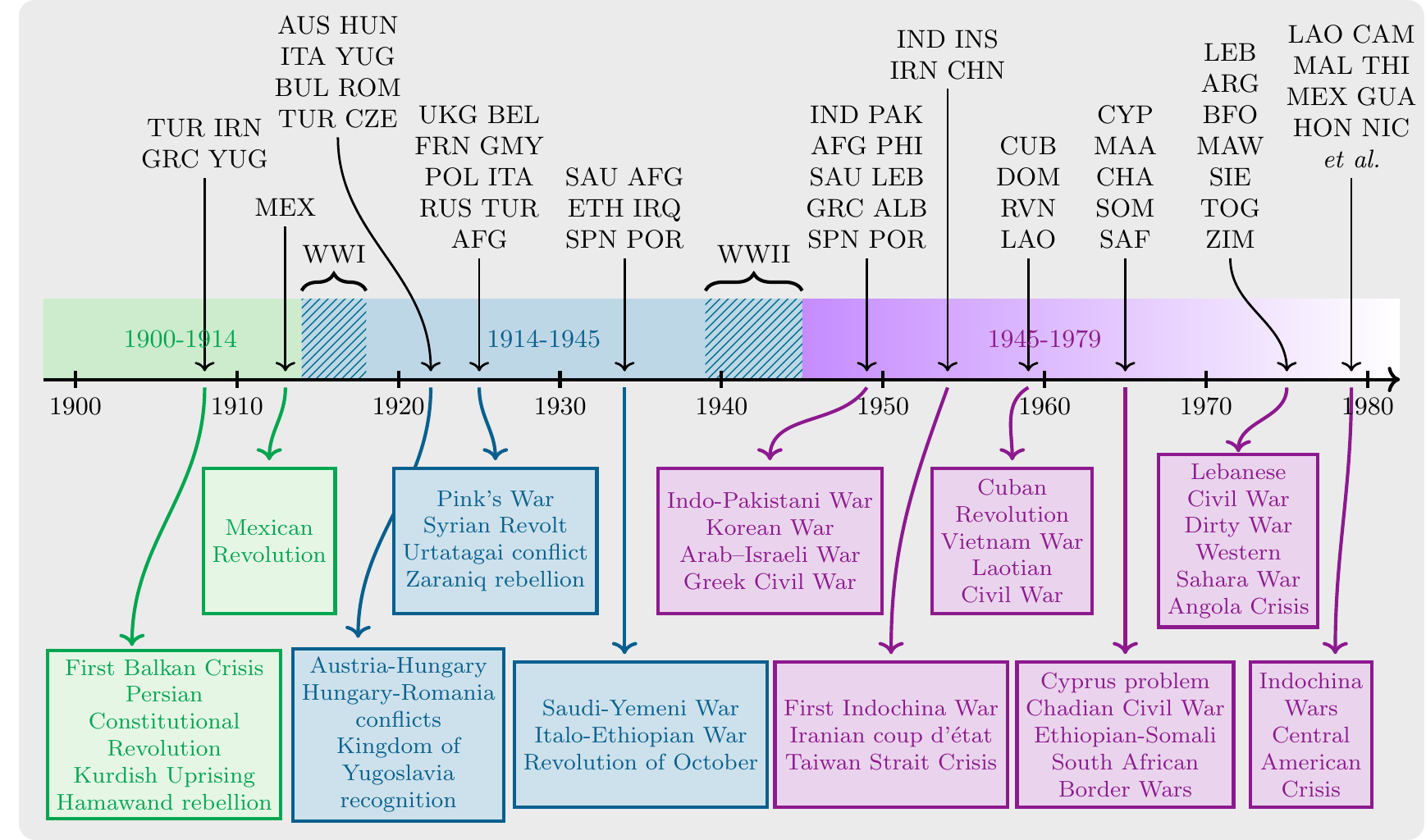}
		\par\end{centering}
	\caption{Timeline of the major events corresponding to valleys detected in
		the XX century. A table relating each country to its corresponding
		code is provided in the SM.}
	\label{timeline2}
\end{figure}

\begin{figure}[H]
	\begin{centering}
		\includegraphics[width=0.9\textwidth]{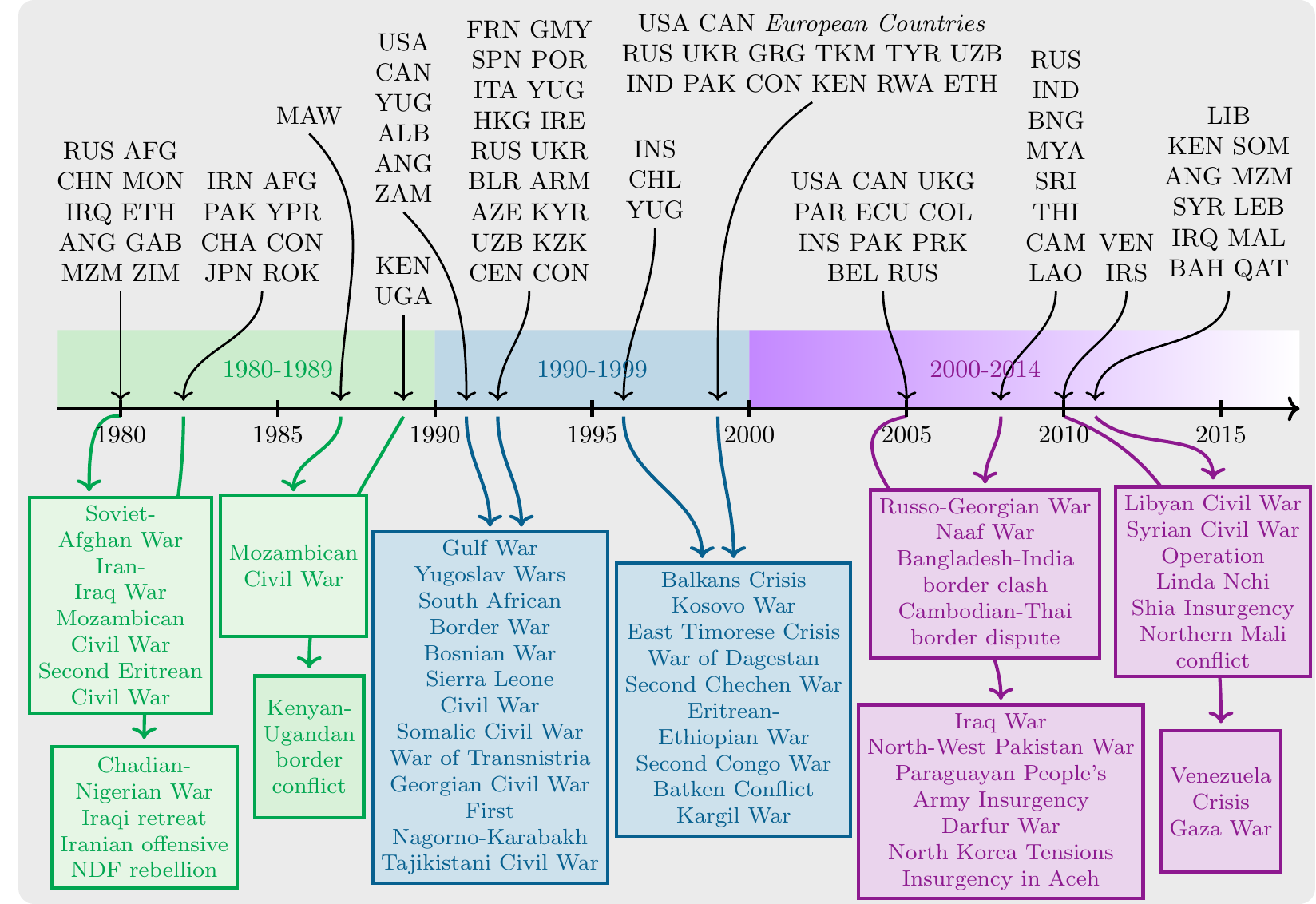}
		\par\end{centering}
	\caption{Timeline of the major events corresponding to valleys detected in
		the end of the XX century and beginning of the XXI century. A table
		relating each country to its corresponding code is provided in the SM.}
	\label{timeline3}
\end{figure}

\begin{figure}[H]
	\begin{centering}
		\subfloat[]{\includegraphics[width=0.45\textwidth]{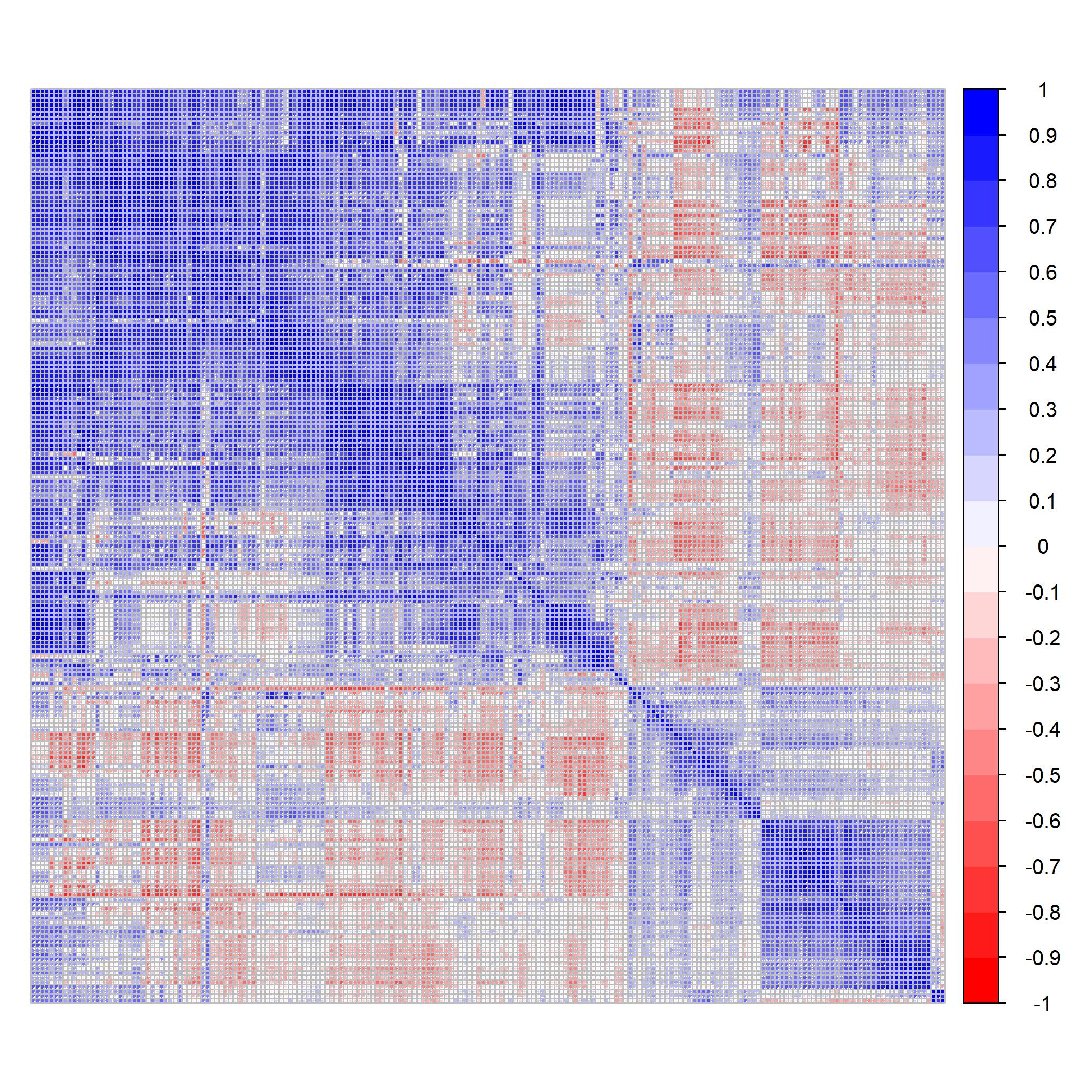}
			
		}\subfloat[]{\includegraphics[width=0.45\textwidth]{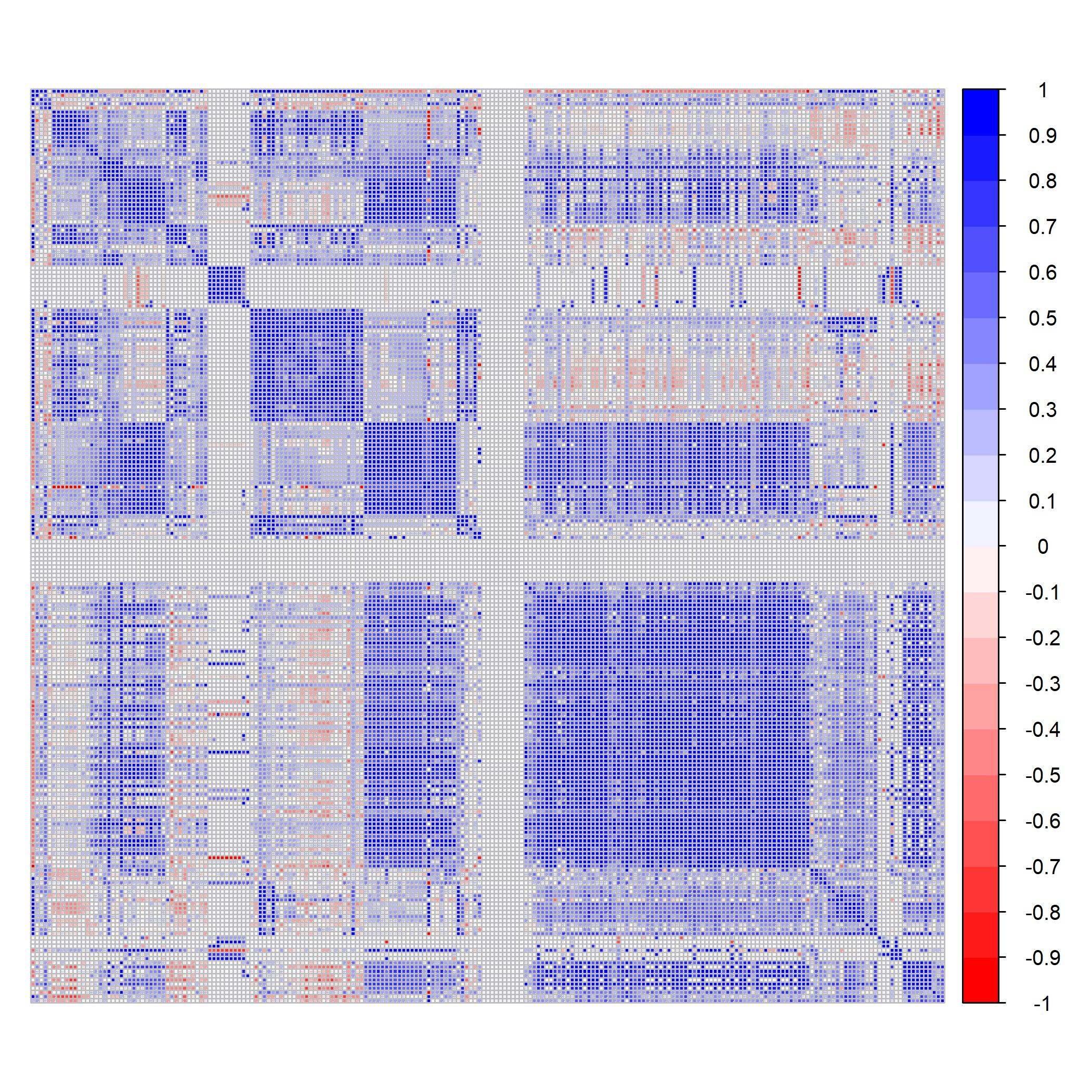}
		}
		\par\end{centering}
	\caption{Pearson correlation plots for the local balance. Panel
		(a): correlogram between the values of the local balance $\kappa_{v}(t)$
		for each pair of years over the full set of countries. Panel (b): correlogram
		between the time series of the local balance $\kappa_{v}(t)$ for
		each pair of countries over the full temporal network from 1816 to 2014.}
	\label{Correlograms}
\end{figure}

\begin{table}[H]
\begin{center}
\begin{tabular}{|c|cc|cc|}
			\hline 
			& \multicolumn{2}{c|}{real} & \multicolumn{2}{c|}{random}\tabularnewline
			\hline 
			Country & Spearman  & p-value & Spearman  & p-value\tabularnewline
			\hline 
			USA & $-0.548$ & $<10^{-9}$ & 0.000486  & 0.52 \tabularnewline
			France & $-0.414$ & $<10^{-5}$ & -0.001147  & 0.51 \tabularnewline
			Portugal & $-0.367$ & $<10^{-4}$ & 0.006732  & 0.46 \tabularnewline
			Russia & $-0.314$ & $0.0006$ & 0.003541  & 0.49 \tabularnewline
			Italy & $-0.237$ & $0.01$ & -0.004927  & 0.52 \tabularnewline
			UK & $-0.185$ & $0.05$ & -0.001555  & 0.50 \tabularnewline
			China & $-0.161$ & $0.09$ & 0.002657  & 0.50 \tabularnewline
			Japan & $-0.109$ & $0.3$ & 0.002705  & 0.50 \tabularnewline
			Turkey & $-0.007$ & $0.9$ & 0.000354  & 0.50\tabularnewline
			\hline 
\end{tabular}
\end{center}
	\caption{Spearman correlation coefficient between the GPR and local balance	index of 9 countries for the real local balance time series (left columns), as well as average correlations for 500 randomizations of the local balance time series (right columns). The p-value of the associated independence hypothesis is also reported. }
	\label{GPR}
\end{table}

\begin{table}[H]
	\begin{center}
	\begin{tabular}{||c|c|c|c||}
		\hline 
		\textbf{Year} & \textbf{Country} & \textbf{$\Delta\kappa_{v}$} & \textbf{Event}\tabularnewline
		\hline 
		1848 & France & -0.138 & European Revolutions of 1848\tabularnewline
		1854 & Ottoman Empire & -0.143 & Crimean War\tabularnewline
		1864 & Denmark & -0.271 & Second Schleswig War\tabularnewline
		1916 & Mexico & -0.422 & Mexican Revolution\tabularnewline
		1914 & France & -0.259 & First World War\tabularnewline
		1921 & Greece & -0.175 & Greco-Turkish War\tabularnewline
		1934 & Yemen & -0.228 & Saudi-Yemeni War\tabularnewline
		1939 & Finland & -0.132 & Winter War\tabularnewline
		1956 & Vietnam & -0.236 & Vietnam War\tabularnewline
		1991 & USA & -0.215 & Gulf War\tabularnewline
		1999 & Yugoslavia & -0.200 & Kosovo War and Balkan Crisis\tabularnewline
		2010 & Venezuela & -0.394 & Venezuela 'Economic War'\tabularnewline
		\hline 
	\end{tabular}
	\end{center}
	\caption{Examples of historic events corresponding to valleys. }
	\label{table_valleys}
\end{table}

\begin{table}[H]
\begin{center}
	\begin{tabular}{||c|c|c|c||}
		\hline 
		\textbf{Year} & \textbf{Country} & \textbf{$\Delta\kappa_{v}$} & \textbf{Event}\tabularnewline
		\hline 
		1829 & Greece & +0.120 & Independence of Greece\tabularnewline
		1849 & Italy & +0.386 & Peace of Milan and Frankfurt Constitution\tabularnewline
		1866 & Austria-Hungary & +0.407 & Armistice of Nikolsburg and Peace of Prague\tabularnewline
		1907 & Japan & +0.149 & Franco-Japanese Treaty and Japan-Korea Treaty\tabularnewline
		1913 & Persia & +0.384 & Istanbul Protocol\tabularnewline
		1919 & France & +0.239 & Treaty of Versailles and Treaty of Saint-Germain-en-Laye\tabularnewline
		1946 & United Kingdom & +0.101 & Anglo-Thai Peace Treaty\tabularnewline
		1963 & Nigeria & +0.417 & First Nigerian Republic\tabularnewline
		1975 & South Africa & +0.435 & Israel-South Africa Agreement\tabularnewline
		1979 & South Africa & +0.249 & The Southern Africa Non-aggression Pact\tabularnewline
		1991 & China & +0.349 & Sino-Soviet Border Agreement\tabularnewline
		2009 & Turkey & +0.114 & Zurich Protocols\tabularnewline
		\hline 
	\end{tabular}
	\end{center}
	\caption{Examples of historic events corresponding to peaks.}
	\label{table_peaks}
\end{table}

\begin{table}[H]
	\begin{center}
	\begin{tabular}{||l|l||l|l||l|l||}
			\hline
			\multicolumn{6}{|c|}{\bf Country Codes}\tabularnewline\hline
			AAB & Antigua \& Barbuda &
			AFG	& Afghanistan & 
			ALB	& Albania  \tabularnewline \hline
			ALG	& Algeria &
			AND	& Andorra &
			ANG	& Angola  \tabularnewline \hline
			ARG	& Argentina &
			ARM	& Armenia &
			AUH	& Austria-Hungary  \tabularnewline \hline
			AUL	& Australia &
			AUS	& Austria &
			AZE	& Azerbaijan  \tabularnewline \hline
			BAD	& Baden &
			BAH	& Bahrain &
			BAR	& Barbados  \tabularnewline \hline
			BAV	& Bavaria &
			BEL	& Belgium &
			BEN	& Benin  \tabularnewline \hline
			BFO	& Burkina Faso &
			BHM	& Bahamas &
			BHU	& Bhutan  \tabularnewline \hline
			BLR	& Belarus &
			BLZ	& Belize &
			BNG	& Bangladesh  \tabularnewline \hline
			BOL	& Bolivia &
			BOS	& Bosnia and Herzegovina &
			BOT	& Botswana  \tabularnewline \hline
			BRA	& Brazil &
			BRU	& Brunei &
			BUI	& Burundi  \tabularnewline \hline
			BUL	& Bulgaria &
			CAM	& Cambodia &
			CAN	& Canada  \tabularnewline \hline
			CAO	& Cameroon &
			CAP	& Cape Verde &
			CDI	& Ivory Coast  \tabularnewline \hline
			CEN	& Central African Rep. &
			CHA	& Chad &
			CHL	& Chile  \tabularnewline \hline
			CHN	& China &
			COL	& Colombia &
			COM	& Comoros  \tabularnewline \hline
			CON	& Congo &
			COS	& Costa Rica &
			CRO	& Croatia  \tabularnewline \hline
			CUB	& Cuba &
			CYP	& Cyprus &
			CZE	& Czechoslovakia  \tabularnewline \hline
			CZR	& Czech Rep. &
			DEN	& Denmark &
			DJI	& Djibouti  \tabularnewline \hline
			DMA	& Dominica &
			DOM	& Dominican Rep. &
			DRC	& Dem. Rep. of the Congo  \tabularnewline \hline
			DRV	& Vietnam &
			ECU	& Ecuador &
			EGY	& Egypt  \tabularnewline \hline
			EQG	& Equatorial Guinea &
			ERI	& Eritrea &
			EST	& Estonia  \tabularnewline \hline
			ETH	& Ethiopia &
			ETM	& East Timor &
			FIJ	& Fiji  \tabularnewline \hline
			FIN	& Finland &
			FRN	& France &
			FSM	& Fed. States of Micronesia  \tabularnewline \hline
			GAB	& Gabon &
			GAM	& Gambia &
			GDR	& German Dem. Rep.  \tabularnewline \hline
			GFR	& German Federal Rep. &
			GHA	& Ghana &
			GMY	& Germany  \tabularnewline \hline
			GNB	& Guinea-Bissau &
			GRC	& Greece &
			GRG	& Georgia  \tabularnewline \hline
			GRN	& Grenada &
			GUA	& Guatemala &
			GUI	& Guinea  \tabularnewline \hline
			GUY	& Guyana &
			HAI	& Haiti &
			HAN	& Hanover  \tabularnewline \hline
			HON	& Honduras &
			HSE	& Hesse Electoral &
			HSG	& Hesse Grand Ducal  \tabularnewline \hline
			HUN	& Hungary &
			ICE	& Iceland &
			IND	& India  \tabularnewline \hline
			INS	& Indonesia &
			IRE	& Ireland &
			IRN	& Iran  \tabularnewline \hline
			IRQ	& Iraq &
			ISR	& Israel &
			ITA	& Italy  \tabularnewline \hline
			JAM	& Jamaica &
			JOR	& Jordan &
			JPN	& Japan  \tabularnewline \hline
			KEN	& Kenya &
			KIR	& Kiribati &
			KOR	& Korea  \tabularnewline \hline
			KOS	& Kosovo &
			KUW	& Kuwait &
			KYR	& Kyrgyzstan  \tabularnewline \hline
			KZK	& Kazakhstan &
			LAO	& Laos &
			LAT	& Latvia  \tabularnewline \hline
			LBR	& Liberia &
			LEB	& Lebanon &
			LES	& Lesotho  \tabularnewline \hline
			LIB	& Libya &
			LIE	& Liechtenstein &
			LIT	& Lithuania  \tabularnewline \hline
			LUX	& Luxembourg &
			MAA	& Mauritania &
			MAC	& Macedonia  \tabularnewline \hline
			MAD	& Maldives &
			MAG	& Madagascar &
			MAL	& Malaysia  \tabularnewline \hline
			MAS	& Mauritius &
			MAW	& Malawi &
			MEC	& Mecklenburg Schwerin  \tabularnewline \hline
			MEX	& Mexico &
			MLD	& Moldova &
			MLI	& Mali  \tabularnewline \hline
			MLT	& Malta &
			MNC	& Monaco &
			MNG	& Montenegro  \tabularnewline \hline
			MOD	& Modena &
			MON	& Mongolia &
			MOR	& Morocco  \tabularnewline \hline
			MSI	& Marshall Islands &
			MYA	& Myanmar &
			MZM	& Mozambique  \tabularnewline \hline
			NAM	& Namibia &
			NAU	& Nauru &
			NEP	& Nepal  \tabularnewline \hline
			NEW	& New Zealand &
			NIC	& Nicaragua &
			NIG	& Nigeria  \tabularnewline \hline
			NIR	& Niger &
			NOR	& Norway &
			NTH	& Netherlands  \tabularnewline \hline
			OMA	& Oman &
			PAK	& Pakistan &
			PAL	& Palau  \tabularnewline \hline
			PAN	& Panama &
			PAP	& Papal States &
			PAR	& Paraguay  \tabularnewline \hline
			PER	& Peru &
			PHI	& Philippines &
			PMA	& Parma  \tabularnewline \hline
			PNG	& Papua New Guinea &
			POL	& Poland &
			POR	& Portugal  \tabularnewline \hline
			PRK	& North Korea &
			QAT	& Qatar &
			ROK	& South Korea  \tabularnewline \hline
			ROM	& Romania &
			RUS	& Russia &
			RVN	& Rep. of Vietnam  \tabularnewline \hline
			\end{tabular}
		\end{center}
\caption{Country Codes used in the text}
	\label{codes1}
\end{table}	

\begin{table}[H]\ContinuedFloat
	\begin{center}
		\begin{tabular}{||l|l||l|l||l|l||}
			\hline
			\multicolumn{6}{|c|}{\bf Country Codes}\tabularnewline\hline
				RWA	& Rwanda &
			SAF	& South Africa &
			SAL	& El Salvador  \tabularnewline \hline
			SAU	& Saudi Arabia &
			SAX	& Saxony &
			SEN	& Senegal  \tabularnewline \hline
			SEY	& Seychelles &
			SIC	& Two Sicilies &
			SIE	& Sierra Leone  \tabularnewline \hline
			SIN	& Singapore &
			SKN	& St. Kitts and Nevis &
			SLO	& Slovakia  \tabularnewline \hline
			SLU	& St. Lucia &
			SLV	& Slovenia &
			SNM	& San Marino  \tabularnewline \hline
			SOL	& Solomon Islands &
			SOM	& Somalia &
			SPN	& Spain  \tabularnewline \hline
			SRI	& Sri Lanka &
			SSD	& South Sudan &
			STP	& Sao Tome and Principe  \tabularnewline \hline
			SUD	& Sudan &
			SUR	& Suriname &
			SVG	& St. Vincent \& the Grenadines  \tabularnewline \hline
			SWA	& Swaziland &
			SWD	& Sweden &
			SWZ	& Switzerland  \tabularnewline \hline
			SYR	& Syria &
			TAJ	& Tajikistan &
			TAW	& Taiwan  \tabularnewline \hline
			TAZ	& Tanzania &
			THI	& Thailand &
			TKM	& Turkmenistan  \tabularnewline \hline
			TOG	& Togo &
			TON	& Tonga &
			TRI	& Trinidad and Tobago  \tabularnewline \hline
			TUN	& Tunisia &
			TUR	& Turkey &
			TUS	& Tuscany  \tabularnewline \hline
			TUV	& Tuvalu &
			UAE	& United Arab Emirates &
			UGA	& Uganda  \tabularnewline \hline
			UKG	& United Kingdom &
			UKR	& Ukraine &
			URU	& Uruguay  \tabularnewline \hline
			USA	& United States of America &
			UZB	& Uzbekistan &
			VAN	& Vanuatu  \tabularnewline \hline
			VEN	& Venezuela &
			WRT	& Wuerttemburg &
			WSM	& Samoa  \tabularnewline \hline
			YAR	& Yemen Arab Rep. &
			YEM	& Yemen &
			YPR	& Yemen People's Rep.  \tabularnewline \hline
			YUG	& Yugoslavia &
			ZAM	& Zambia &
			ZAN	& Zanzibar  \tabularnewline \hline
			ZIM	& Zimbabwe & - & - & - & -  \tabularnewline \hline
		\end{tabular}
	\end{center}
\caption{Country Codes used in the text}
	\label{codes2}
\end{table}	

{\bf Conflict of interest statement}

Not Applicable.



\end{document}